\documentclass[11pt,twoside]{llncs2}
\usepackage{doublespace}
\usepackage{epic}
\usepackage{note3}
\usepackage{fullpage}

\usepackage{epsfig,amsmath,latexsym,amssymb}
\usepackage{psfig}

\usepackage{verbatim}
\usepackage{color}

\newcommand{\nc}{\newcommand}
\nc{\rnc}{\renewcommand}

\nc{\smfrac}[2]{\mbox{$\frac{#1}{#2}$}}
\nc{\ox}{\otimes}
\nc{\dg}{\dagger}
\def\ot{\otimes}
\nc{\eq}[1]{(\ref{eq:#1})}
\nc{\eqs}[2]{(\ref{eq:#1}) and (\ref{eq:#2})}
\nc{\eqm}[2]{(\ref{eq:#1})--(\ref{eq:#2})}

\rnc{\sec}[1]{Sect$.\,$\ref{sec:#1}}
\nc{\fig}[1]{Fig$.\,$\ref{fig:#1}}

\def\lbm{ \left[\rule{0pt}{2.1ex}\right. }
\def\rbm{ \left.\rule{0pt}{2.1ex}\right] }

\def\lbL{ \left[\rule{0pt}{2.4ex}\right. \!}
\def\rbL{ \!\left.\rule{0pt}{2.4ex}\right] }

\rnc\ss{\hspace*{0.1ex}}
\nc\ms{\hspace*{-0.1ex}}

\def\Pr{{\rm Pr}}
\def\Tr{{\rm Tr}}
\def\be{\begin{eqnarray}}
\def\ee{\end{eqnarray}}
\def\bea{\begin{eqnarray}}
\def\eea{\end{eqnarray}}

\def\ra{\rightarrow}

\def\e{\epsilon}

\def\G{\Gamma}
\def\s{\sigma}
\def\sr{s.r.$\;$}

\def\I{{\rm I}}

\def\P{{\cal P}}
\def\S{{\cal S}}
\def\CM{{\cal M}}
\def\CE{{\cal E}}
\def\CD{{\cal D}}
\def\CI{{\cal I}}
\def\CR{{\cal R}}

\rnc{\AA}{{\mathbb A}}
\nc{\Q}{{\mathbb Q}}
\rnc{\H}{{\mathbb H}}
\nc{\BB}{{\mathbb B}}
\nc{\EE}{{\mathbb E}}
\nc{\XX}{{\mathbb X}}
\nc{\UU}{{\mathbb U}}
\nc{\C}{{\mathbb C}}

\def\>{\rangle}
\def\<{\langle}
\def\ot{{\otimes}}

\nc{\mPr}[1]{
    \begin{array}{c}
    \rule{0pt}{0.0ex}\mbox{\rm Pr} \\ \raisebox{0ex}{\scriptsize $#1$}
    \end{array} \!\!
}
\rnc{\mPr}[1]{
    \raisebox{1.5ex}\mbox{\rm Pr} \hspace{-2ex}
    \raisebox{-1.7ex}{\scriptsize $#1$}~
}

\nc{\mrho}{\raisebox{0.15ex}{$\rho $}}

\newcommand{\acc}{\textsc{acc}}
\newcommand{\rej}{\textsc{rej}}

\rnc{\comment}[1]{{}}

\nc{\hhline}[3]{\put(#1,#2){\line(1,0){#3}}}
\nc{\vvline}[3]{\put(#1,#2){\line(0,1){#3}}}
\nc{\ffbox}[5]{\put(#1,#2){\framebox(#3,#4){#5}}}
\nc{\mmbox}[5]{\put(#1,#2){\makebox(#3,#4){#5}}}

\begin{document}

%---------------------------------------------------------------------
\title{The Universal Composable Security of \\ Quantum {Message}
Authentication with Key Recyling\\[-2ex]}

\titlerunning{The Universal Composable Security of Quantum Message
Authentication with Key Recyling}

\author{Patrick Hayden\inst{1,2}, Debbie W.\ Leung\inst{1,3},
Dominic Mayers\inst{1}}
\institute{Institute for Quantum Information, Caltech, Pasadena,
California, USA
\and
Physics Department,
Stanford University, Stanford, California, USA
\and
Institute for Quantum Computing,
University of Waterloo, Waterloo, Ontario, Canada \\
\vspace*{1ex}
\email{
phayden\@@stanford.edu, wcleung\@@uwaterloo.ca,
dmayers\@@cs.caltech.edu}}

\authorrunning{P.~M.~Hadyen\inst{1,2}, D.~W.~Leung\inst{1,3},
D.~Mayers\inst{1,4}}

\date{\today}

\maketitle

% TQA' checks entanglement first, TQA doesn't. 

\begin{abstract}
Barnum, Cr\'{e}peau, Gottesman, Tapp, and Smith \cite{BCGST02}
proposed methods for authentication of quantum messages.  The first
method is an interactive protocol (TQA') based on teleportation.  The
second method is a noninteractive protocol (QA) in which the sender
first encrypts the message using a protocol QEnc
and then encodes the quantum ciphertext with an error correcting code
chosen secretly from a set (a purity test code (PTC)).   
Encryption was shown to be necessary for authentication.

We augment the protocol QA with an extra step which recycles the entire
encryption key provided QA accepts the message.  
We analyze the resulting integrated protocol for quantum authentication 
and key generation, which we call QA+KG.
Our main result is a proof that QA+KG is universal composably (UC)
secure in the Ben{\hspace{-0.2ex}-\hspace{-0.2ex}}Or--Mayers model
\cite{BM02}.
More specifically, this implies the UC-security of (a) QA, (b)
recycling of the encryption key in QA, and (c) key-recycling of 
the encryption scheme QEnc
by appending PTC.  
For an $m$-qubit message, encryption requires $2m$ bits of key; but PTC 
can be performed using only $O(\log m) + O(\log \epsilon)$ bits of key for 
probability of failure $\epsilon$.   
Thus, we reduce the key required for both QA and QEnc, from linear to
logarithmic net consumption, at the expense of one bit of back
communication which can happen any time after the conclusion of QA and
before reusing the key.
UC-security of QA also extends security to settings not obvious from 
\cite{BCGST02}.

Our security proof structure is inspired by and similar to that of
\cite{BCGST02}, reducing the security of QA to that of TQA'.  In the
process, we define UC-secure entanglement, and prove the
UC-security of the entanglement generating protocol given in
\cite{BCGST02}, which could be of independent interest.
\end{abstract}

\setstretch{0}
\setlength{\parskip}{1ex}
\setlength{\parindent}{0ex}

%\raggedbottom
%\thispagestyle{empty}

%%%%%%%%%%%%%%%%%%%%%%%%%%%%%%%%%%%%%%%%%%%%%%%%%%%%%%%%%%%%%%%%%%%%%
\section{Context, results and related work}
\label{sec:motivation}

{\bf Encryption and authentication of quantum messages}

Barnum, Cr\'{e}peau, Gottesman, Tapp, and Smith \cite{BCGST02} studied
authentication of quantum messages.  Their first proposed method is an
interactive protocol (TQA') based on teleportation.  Entanglement is
first established between the sender, Alice, and the receiver, Bob,
via an insecure quantum channel, using a method called the purity test
protocol (PTP).  If that is successful, the quantum message is
teleported.  A two-way authenticated classical channel is assumed.
Their second proposed method is a noninteractive protocol (QA) in
which the sender first encrypts the message (using a protocol called
QEnc \cite{AMTW00,Boykin00} and $2m$ bits of key for an $m$-qubit
message) and then encodes the quantum ciphertext with an error
correcting code chosen secretly from a set.  QA rejects/accepts if an
error is/not detected.  The set of possible error correcting codes is
called the purity test code (PTC).  Each code uses $s = s' + 1/2 -
\log(m/s)$ extra qubits of communication, and takes $2s'+ 2 + 2
\log(m/s')$ key-bits to choose secretly, in order to achieve a
probability of failure (as defined below) of $\epsilon \leq 2^{-s'}$.
Unlike authentication of classical messages, which can be done without
encryption and with key size $O(\log(m))$, \cite{BCGST02} proved the
necessity of encryption in quantum authentication. Thus, in the
noninteractive setting, the key length must be at least $2m$ required
for encryption \cite{AMTW00}.

% which uses only a key logarithmic in the security
% parameter.
%
% true only if s is at least log m
% in fact, if security parameter is 2^{-s'},
% s' \approx s-1-log(m/s), the key cost = 2s+1 = 2s'+ 2 + 2 log(m/s)

\clearpage

{\bf Key recycling -- intuition and early ideas}

The protocol QA is somewhat analogous to the classical scheme due to
Wegman and Carter \cite{Wegman81}.  The latter requires a large key
but most of it can be reused so that only a logarithmic sized key is
actually consumed.
A natural question is whether it is possible to reuse part of the key
required in QA.
For quantum messages, successful eavesdropping necessarily causes 
disturbance \cite{Bennett94a}.  
This insight \cite{BBB82,BBBS84}, which even then suggested the possibility 
of key recycling, led to the original discovery of
quantum key distribution (QKD) \cite{BB84}.
In \cite{Leung00}, encryption using a quantum key with
recycling was proven secure, and the question arose whether the
classical key in QEnc could likewise be securely recycled.  Using
two-way classical discussion to implement a form of quantum
authentication before it was formalized in \cite{BCGST02}, some 
security statements were obtained. 
Qualitatively, it is unlikely for a quantum message to be
authenticated and accepted if it has been attacked and the key been
compromised.  This opens the tantalizing possibility of reusing the key
whenever the message is accepted.
However, a proper security statement for key recycling can be hard to
formulate, let alone be obtained, because it requires an analysis of the
most general {\em joint quantum} attack on quantum authentication
together with the scheme that subsequently uses the recycled key.
To complicate matters further, the usual security measure for a key in
terms of Eve's classical mutual information was found to be highly
unstable with respect to additional classical information on the key
(see \cite{DiVincenzo03}, the motivation for \cite{BHLMO-TCC05} and
\cite{RK05}).

\vspace*{2ex}

{\bf The universal composability approach}

To resolve these questions in a robust way, we analyze the security of
QA and key recycling in the framework of universal composability.
This also proves that QA has some additional nontrivial security
features.

Composability is concerned with the security of composing
cryptographic primitives in a possibly complex manner.
The simplest example is the security of using a cryptographic
primitive as a subroutine in another application.
We will follow the {\em universal composability (UC)} approach:
For a specific task (functionality), a primitive that realizes the
task is defined to be universally composable if it cannot be
distinguished (up to a bias which is the security parameter) from the
ideal functionality (augmented with a simulator) by any
``environment'' that controls the input, retains a purification, provides 
it to the adversary, 
directs the adversarial attacks, and receives the state possessed by the
adversary as well as all user outputs of the protocol.
Any application using the primitive (as a subroutine) is provably essentially
as secure as one using the ideal functionality.  Also, a recursive
argument for security holds for a composite protocol with acyclic modular
structure, and the resulting security parameter is at most additive.
A security definition that ensures {\em universal} composability was
recently proposed by Canetti \cite{Canetti01} in the classical
setting.
A simpler model in the quantum setting and a corresponding universal
composable security definition were reported in \cite{QIP03-1,BM02}.
(An alternative approach to composability was obtained in \cite{BPW04}
in the classical setting, and generalized to the quantum setting in
\cite{Unruh04,Unruh04b}.)

Since we are concerned with unconditional security, the analysis is
particularly simple -- it suffices to show that the actual primitive
and the ideal functionality (augmented with a simulator) cannot be
distinguished by any physical process.
Universal composability provides a systematic, general and robust
framework for analyzing the security of recycled key, even in the
presence of subtleties including entanglement and collective attacks.

\clearpage

{\bf Our techniques, proof structure, and results}

In \cite{BCGST02}, security for quantum authentication is defined in
terms of the probability of failing to reject in the presence of a
detectable error.  The authors consider a protocol TQA which is
similar to TQA' except Bob never tells Alice whether the entanglement
is accepted or rejected.  In TQA, the purity test code (PTC) is only
used in a subroutine to establish entanglement (ebits) between Alice
and Bob.  The authors proved the security of TQA and the equivalence
of the security of QA and TQA.

To analyze the security of key recycling in the UC framework, we
consider an augmented protocol, QA+KG, which recycles the $2m$
key-bits used in the QEnc step if QA accepts.  
Note that key recycling requires Alice to know whether QA accepts or
rejects.
We model our ideal functionality for key generation for the
non-interactive protocol QA+KG, such that if Alice further receives one
bit of back communication from Bob, she can complete the ideal key
recycling step.
With this in mind: 

(1) We show QA+KG and TQA+KG are indistinguishable to any environment.
Thus the two protocols QA and TQA still have equivalent securities
even with key recycling and in the UC framework.

We also formalize how TQA uses a subroutine ``EBIT[PTC]'' which
generates entanglement via insecure channel using PTC as a subroutine.
TQA+KG teleports \cite{TP} the quantum message using EBIT[PTC] and a
perfect encrypted and authenticated classical channel denoted by
C$_\I$.  After using the classical message to complete teleportation,
it is output as a key.  In other words, the protocol TQA+KG can be
interpreted as (TP+KG)[EBIT[PTC],C$_\I$] where TP stands for
teleportation.

(2) Following \cite{BCGST02}, and applying results from there, we show
that EBIT[PTC] cannot be distinguished from a different protocol
EBIT[PTP] for generating entanglement.

(3) We show that EBIT[PTP] and the ideal functionality EBIT$_\I$ for
generating entanglement cannot be distinguished by any environment
with bias better than $2 \sqrt{2} \epsilon^{1/3}$ where $\epsilon$ is
the probability of failure in PTP.  This proves that EBIT[PTP] is a UC
secure method to generate entanglement.  

Technically, this is the only step in our proof that involves the
construction of a ``simulator'' which is crucial in directly
establishing the UC security of a protocol.  (The rest of our proof 
relies on transitivity and the composability theorem.)  

(4) We apply (2)-(3) and the composability theorem to show that
TP[EBIT[PTC],C$_\I$] and the ideal channel Q$_\I$=TP[EBIT$_\I$,C$_\I$]
cannot be distinguished with bias great than $2 \sqrt{2} \epsilon^{1/3}$.

(5) Finally, we show that (TP+KG)[EBIT[PTC],C$_\I$] and Q$_\I$+KD$_\I$
(where KD$_\I$ denotes an ideal key generating functionality) cannot
be distinguished by any environment with bias greater than $2 \sqrt{2}
\epsilon^{1/3}$.  The intuition is that, replacing EBIT[PTC] by
EBIT$_\I$ in (TP+KG)[EBIT[PTC],C$_\I$] also protects the 
classical teleportation message which then can be reused as a key.

Together, QA+KG is distinguishable from Q$_\I$+KD$_\I$ with bias at 
most $2 \sqrt{2} \epsilon^{1/3}$.  

We thus prove that QA+KG is UC secure in the Ben-Or-Mayers model
\cite{BM02}.
More specifically, this implies the UC-security of (a) QA, (b)
recycling of the encryption key in QA, and (c) key-recycling of QEnc
by appending PTC.  
We reduce the key required for both QA and QEnc, from linear in the
message size to a logarithmic net consumption (if QA accepts), at the
expense of one bit of back communication which can happen any time
after the conclusion of QA and before using the recycled key.
Furthermore, UC-security of QA implies it can be used securely in
other cryptographic tasks.  In particular, parallel composition is
secure against joint attacks, and QA is still secure if the
adversary possesses the purifying system of the message to be
authenticated.  These are not immediate consequences of the analysis
in \cite{BCGST02}.

In the process, we define UC security for entanglement generation and
prove the UC-security for the protocol EBIT[PTC] proposed in
\cite{BCGST02}, which is of independent interest.
% We prove UC-security for the Wegman-Carter
% scheme \cite{Wegman81} against quantum adversaries, removing the need
% of authenticated classical channel.

Our result does not contradict earlier lower bounds on the key size,
which applies to noninteractive protocols and is concerned with the
initial key needed.  
Another nice aspect of our results is that one can simply reuse the
encryption key without further privacy amplification, in contrast
to quantum key distribution and earlier proposals for key
recycling.

\vspace*{2ex}

{\bf Prior and related work}

We have discussed background results leading to this investigation
(which started 2003) -- the QEnc and QA protocols in
\cite{AMTW00,BCGST02}, UC security \cite{BM02} and some early
investigations of key recycling \cite{Leung00}.  Our proof steps are
similar to those in \cite{BCGST02}, but we resolve definitional
ambiguities in \cite{BCGST02} and with the composability framework
obtain more precise and stronger security results (UC security of QA,
QEnc (by adding PTC), EBIT[PTC], EBIT[PTP], and key recycling).
Throughout, we may emphasize the modular structure of a protocol
${\cal P}$ calling a subroutine $\sigma$ by writing ${\cal
  P}[\sigma]$.

We now discuss other related works since this project started.  

The security of key recycling in QA was studied independently by
M.~Horodecki and Oppenheim~\cite{HO03} in 2003.  However \cite{HO03}
does not address the security of QA, and it assumes an adversary who does
not possess the purification. %(end of p4, arxiv version).  
For that reason, we believe their claim to UC security, even if it holds,
requires a nontrivial proof, but none was given.

In 2005, Damgard, Pedersen, and Salvail \cite{BRICS05} proposed key
recycling for the encryption of {\em classical messages} by using the
Wegman-Carter classical authentication scheme followed by a quantum
encryption scheme based on key uncertainty or locking
\cite{DiVincenzo03}.
Encryption of quantum messages was said to be possible in the
introduction but no proof of this assertion was given in the text.
Regardless, the results in \cite{BRICS05} are quite different from
ours because encryption and authentication of classical messages are
much weaker tasks cryptographically.
Also, locking is highly non-composable when a quantum adversary has
quantum memory and delays measurements.  It is unclear how the
analysis in \cite{BRICS05} fits into the composability framework,
despite a claim (without formal definition or proof) of the composable
security of the regenerated key.
(We detail the differences here since an earlier version of our paper
was rejected in 2007 by a referee who assumed this work to be similar
to \cite{BRICS05}.)

In this paper, we emphasize the necessity of considering the
composable security of the regenerated key, since the entire purpose
of recycling the key is to use it later.  Furthermore, universal
composability is precisely what allows the key to be used in a
yet-to-be-specified and unrestricted manner. Without such an
assurance, the security of key recycling is ill-defined.

This paper has had an unusually long gestation. We presented a
preliminary version of our results at QIP 2004 and a draft has
informally circulated since 2008. An updated version appeared in
QCRYPT 2011.  (The full submission was provided to the authors of
related works \cite{P13,GYZ16,P16} prior to their appearing in 
the arXiv.)

Until this year, ours remained the only proof of the composable
security of QA and of key recycling in QA and QEnc. A flurry of recent
activity in the area by other authors prompted us to produce and
submit this final version of our draft.

First, Garg, Yuen, and Zhandry \cite{GYZ16} gave a new definition of
quantum authentication called ``total authentication'' that they
showed to permit composably secure key recycling.  They further
exhibited several new protocols satisfying the definition.  Our work
implied that QA proposed in \cite{BCGST02} satisfies total
authentication but with a very small key leakage.

More recently, Portmann \cite{P16} has established proofs of both of
our main results in the framework of abstract cryptography \cite{MauRen11}.
Furthermore, partial key recycling is proven secure when
authentication fails.
His work formally considers impersonation attack, whereas all other work
implicitly assumes this is a special case of the substitution attack.
His work also explicitly considers communication over noisy channels.  We
have added a short discussion to our paper that illustrates how secure
authentication (with key recycling) over noisy channels is an
immediate corollary of composable security of QA+KG.

One feature that slightly distinguishes \cite{GYZ16,P16} from ours is
that they demonstrate that the entire key can be recycled whereas we
sacrifice a vanishing fraction of the key.  While interesting
theoretically, the distinction is not practically important because,
in our case, additional key to make up for the small loss can be added
to the message with negligible additional cost.  Furthermore, some
of the schemes that allow total key recycling require substantially
more initial key (while QA is key-optimal up to an additive
logarithmic amount, which we believe can be reduced to a constant 
in view of results in \cite{aharonov2014local}).  

In another recent contribution, somewhat closing the circle, Fehr and
Salvail \cite{FS16} proved that secret key could be securely recycled
in a quantum protocol for authenticating classical messages. Their
protocol is a slightly modified version of one first proposed by
Bennett, Brassard, and Breidbart in 1982 \cite{BBB82} and within reach
using current experimental techniques.

The current manuscript differs from our QCRYPT'11 submission in four
ways.  (1) We found a mis-statement of the adversarial power in the
QCRYPT'11 submission which is corrected here -- the adversary should
be given the purification of the message (full quantum side
information) for the attack.  Our proof is independent of whether the
adversary is given this purification or not.  (2) We simplified the
last step of the proof (and as a bonus reduced the insecurity
parameter by a factor of 3).  (3) In view of \cite{P13}, we removed
claims of proof of the composability of the Wegman-Carter
authentication scheme for classical messages in this paper.  Our claim
was based on a simple (but slightly mistaken) proof in a half-page
appendix.  We decide to keep the appendix for readers who want a quick
main idea, but refer to the detailed subsequent result in \cite{P13}.
(4) We revived an appendix on authentication of pure quantum states
(which was removed in QCRYPT'11 due to page limit).  Finally, as
mentioned earlier, we briefly discussed the case for transmission
through noisy channel, and made other minor changes.

% Earlier versions of our result was presented in 2004 and a draft has
% informally circulated since 2008.  
% 
% Our result was subsequently accepted for presentation in QCRYPT'11,
% and our QCRYPT'11 full submission was further circulation.
% c16 (see following discussion).
% 

\vspace*{2ex}

{\bf Comparison with other methods}

There are two other quantum methods that provide similar security for
quantum message encryption and authentication using only a small key.
We now compare other costs, such as the amount of (forward) quantum
communication, forward and backward classical communication, and the
round/interaction complexity.

If the original message has $m$ qubits, QA+KG consumes a little more
than $m$ qubits of communication, one bit of back communication (which
can be delayed until right before reusing the key), and a little more
than $2m$ key-bits; $2m$ of them can be regenerated if
authentication accepts.

A first alternative to this approach is to use QKD to securely expand
the classical key before running QA without key recycling.  This
requires only a small initial key (not just the amortized one).  The
drawback is that the QKD protocol itself needs at least $2m$ qubits, two
rounds of classical back communication, and a linear amount of forward
classical communication.  The first round of back communication is to
acknowledge the receipt of the quantum states by Bob, followed by $1$-
or $2$-way public discussion (that itself has to be authenticated) and
finally more back communication to finalize the output key size.
Then, $m$ more qubits have to be sent in QA.
Thus, this method consumes substantially more quantum communication
and forward classical communication, and one extra round of back
communication.  Furthermore, the back communication has to be performed during
the protocol.
Running QA before QKD requires the long $2m$-bit initial key, but the
back communication can be delayed until the QKD is run (but before
the application using the key produced).

A second alternative is TQA -- teleport the quantum message using
ebits obtained by potentially insecure means in addition to an
insecure forward classical channel that needs classical message
authentication.
(Since we prove the composable security of EBIT[PTC] and the
Wegman-Carter scheme is composably secure \cite{P13}, this method 
is composably secure.)
Classical message authentication requires a long key, but most of it
can be reused securely regardless of the authentication result.
EBIT[PTC] uses a small key and back communication, and generates a
quantum key. (Thus, back communication is needed during the protocol
itself, unlike for QA+KG.)
Compared to QA+KG, this scheme uses a similar amount of quantum
communication, more initial key and forward classical communication,
in addition to a similar amount of classical back communication.
But TQA offers two advantages over QA+KG. First, failing PTC when
generating ebits does not destroy the quantum message itself (so the
message is not only authenticated, but protected).  Second, the 
classical authentication key is
always recycled.

We emphasize that these methods and QA+KG are incomparable and
interesting for different reasons.  Also, we are concerned not only
with the key requirement, but security definitions and composability
of protocols like QA and EBIT[PTC].

\vspace*{2ex}

{\bf Organization of the paper}

We will discuss background materials concerning the security setting,
quantum mechanics, quantum encryption, quantum authentication, and
quantum universal composability in \sec{basics}, and prove the
security of QA+KG in \sec{proof}.  Other results and open questions
will be discussed in \sec{discussion}.  A glossary, the quantum UC
model, the extended transpose trick, a simple (but slightly mistaken)
proof of the quantum UC-security of the Wegman-Carter scheme, and the
security of authentication for pure quantum states with half of the
initial key cost are given in the appendices.

\clearpage
%%%%%%%%%%%%%%%%%%%%%%%%%%%%%%%%%%%%%%%%%%%%%%%%%%%%%%%%%%%%%%%%%%%%%
\section{Setting, notation, definitions, and background}
\label{sec:basics}

% \vspace*{-3ex}

%-----------------------------------------------------------------------
% \subsection{Preliminary materials}

{\bf Setting.~~} The sender, the receiver, and the adversary are
called Alice, Bob, and Eve, respectively.  We consider unconditional
security, i.e., security against an Eve whose capabilities are only
limited by quantum mechanics.

{\bf Basic elements of quantum mechanics.~~}
A quantum system with $d$ distinguishable states is associated with
the $d$-dimensional complex Hilbert space $\C^d$.  The set of linear
and unitary operators acting on $\C^d$ are denoted by ${\cal B}(\C^d)$
and $\UU(\C^d)$ respectively.  Composite systems are associated with
tensor product Hilbert spaces.

The state of a quantum system is represented by a positive
semidefinite {\em density matrix} $\mrho \in {\cal B}(\C^d)$ of unit
trace.  It is a convex combination (or probabilistic mixture) of
rank-$1$ projectors (commonly called {\em pure states}).
Pure states can be represented as unit vectors $|\psi\> \in \C^d$, up
to a physically unobservable phase, and we write the corresponding
density matrix $|\psi\>\<\psi|$ simply as $\psi$.  Throughout, we
denote an ebit by $|\Phi\> = \smfrac{1}{\sqrt{2}} ( |00\>{+}|11\>)$
and its density matrix by $\Phi$.

A measurement $\CM$ is specified by a POVM --- a set of positive
semidefinite operators $\{O_k\}$ such that $\sum_k O_k = I$.  If the
state is initially $\mrho$, the measurement $\CM$ yields the outcome
$k$ with probability $\Tr(O_k \mrho)$ and changes the state to
$\sqrt{O_k} \mrho \sqrt{O_k}/ \Tr(O_k \mrho)$, without loss of
generality.  $\CM$ is said to be along a basis $\{|k\>\}$ if $\{O_k\}
= \{|k\>\<k|\}$.  Measuring an unknown state generally disturbs it.

The most general evolution of a state is given by a trace-preserving
completely-positive (TCP) linear map $\CE$ acting on ${\cal B}(\C)$.
(See \cite{Nielsen00bk} for various representations.)
Discarding a (sub)system is given by the (partial) trace operation.
Every state $\mrho \in {\cal B}(\C^d)$ can be written as the partial
trace of some pure state $|\psi\> \in \C^d \ot \C^{d'}$.  In other
words, $\mrho = \Tr_2 (\psi)$ and $|\psi\>$ is called its {\em
purification}, and the extra system is called the {\em purifying
system}. 

Subscripts of states and operations often (though not always) 
label the system being acted on.

We mention two distance measures for quantum states.  The first
measure is the trace distance $\frac{1}{2} \| \mrho_1-\mrho_2 \|_1$
between two density matrices $\mrho_1$ and $\mrho_2$, where $\|\cdot
\|_1$ denotes the Schatten $1$-norm.  The maximum probability of
distinguishing the two states drawn randomly is given by
$\frac{1}{2}+\frac{1}{4}\| \mrho_1-\mrho_2 \|_1$.
The second measure is the fidelity, $F(\mrho_1, \mrho_2) =
\max_{|\psi_1\>,|\psi_2\>} |\<\psi_1|\psi_2\>|^2$, where $\mrho_{1,2}
\in {\cal B}(\C)$, $|\psi_{1,2}\> \in \C \otimes \C'$ are
purifications of $\mrho_{1,2}$ and $\<\cdot|\cdot\>$ is the inner
product in $\C \otimes \C'$.  Note that we have an additional square
in the fidelity compared to other references such as
\cite{Nielsen00bk}.

We denote by $\s_{10} {=} \left( \begin{array}{cc} 0 & 1 \\ 1 & 0
\end{array} \right)$,
$\s_{01} {=} \left( \begin{array}{cc} 1 & 0 \\ 0 & {-}1
\end{array} \right)$, and $\s_{11} {=} \s_{10} \s_{01}$
the Pauli matrices acting on $1$ qubit.
The Pauli group acting on $m$ qubits is generated multiplicatively by
$\s_{10}, \s_{01}$ acting on each qubit.

The interested reader can consult \cite{Nielsen00bk} for a more
comprehensive review.

%-----------------------------------------------------------------------
\subsection{Quantum encryption}

\label{sec:qe}

{\bf Definition}

The cryptographic task of quantum encryption can be described as
follows.
Alice and Bob share a key $K$ in which the realization $k$ occurs with
probability $p_k$.  To send a message $\mrho$, Alice transmits
$\CE_k(\mrho)$ and Bob applies $\CD_k$ to retrieve $\mrho$.  A 
quantum encryption scheme should satisfy two properties: 

\emph{Completeness:} $\forall k ~ \CD_k \, \CE_k = \CI$, the identity
operation.

\vspace*{-0.5ex}

\emph{Soundness:} $\CR(\mrho) := \sum_k p_k \, \CE_k(\mrho) = \mrho_0$,
where $\mrho_0$ is a constant.

\vspace*{0.5ex}

The soundness condition is an exact security statement that, without
knowledge of the key, a specimen of the encrypted message $\CR(\mrho)$
is independent of the actual message $\mrho$.
If the message system $M$ is entangled with other systems, let $R$ be
its purifying system, and $|\psi\>_{MR}$ the purification.  By
linearity $(\CR_M \otimes {\cal I}_R) (\psi) = \mrho_0 \otimes
\Tr_M (\psi)$ which means that the transmission is still completely
useless to the strongest eavesdropping adversary who already possesses
all the correlations with $M$ contained in $R$.
A natural approximate security condition is,
$\forall |\psi\>_{MR}, \| (\CR \otimes {\cal I}) (\psi) - \mrho_0
\otimes \Tr_M (\psi) \|_1 < \epsilon$, a small security parameter.
Note $\| \CR(\mrho) - \mrho_0 \|_1 < \epsilon$ is generally too weak
for a security definition \cite{HLSW03}, unless the adversary is
restricted to not having the purifying system $R$.  A scheme that
satisfies this last condition will be called an approximate encryption
scheme (with security parameter $\epsilon$). % $\approx$QEnc.

{\bf Known constructions}

A special case is $\CE_k(\mrho) = U_k \mrho U_k^\dagger$ with each
$U_k$ unitary.
In particular, exact encryption can be achieved by taking $K$ to be a
random $2m$-bit string, and for $k = (x_1,z_1,\cdots,x_m,z_m)$, $U_k =
\s_{x_1 z_1} \otimes \cdots \s_{x_m z_m} =: \s_{xz}$.  We call
this specific protocol QEnc.  There are approximate encryption schemes 
with
certain $U_k$ and $K$ of size $m{+}O(\log m){+}2 \log(1/\epsilon)$ bits
\cite{HLSW03,AS04,DN05} (improved to $m{+}2 \log(1/\epsilon){+}8$ bits
in \cite{A08}).  We focus on the scheme in \cite{HLSW03,A08} and call
it $\approx$QEnc.

{\bf Relation of QEnc to teleportation and remote state preparation,
and lower bounds}

In teleportation (TP) \cite{TP} of $1$ qubit, Alice and Bob share one
ebit in systems $A$ and $B$.  The message $\mrho$ in system $M$ is
transmitted by Alice measuring $MA$ in the Bell basis $\{I \ox
\s_{xz} |\Phi\> \}_{x,z}$.  Conditioned on the outcome $x,z$,
the state in system $B$ is $\s_{xz} \,\mrho \,\s_{xz} $.  Thus, if Bob
knows $x,z$ (sent to him from Alice by a classical channel), he can
recover the message $\mrho$.  An $m$-qubit message can be teleported
qubit-wise.

Thus, there is a one-to-one correspondence between the protocols for 
QEnc and TP; the key
in QEnc translates to the measurement outcome, and thus the
communicated message, in TP.  A similar 
correspondence holds between  
any quantum encryption scheme and a generalized teleportation protocol  
that sends quantum states using classical communication and 
entanglement \cite{AMTW00,LS02}.
Composing generalized teleportation with
superdense coding \cite{SD} to transmit $2m$ classical bits proves
that $2m$ bits is a lower bound on the communication cost of generalized 
teleportation and
thus a lower bound on the key cost in any exact encryption as well.

Likewise, there is a one-to-one correspondence between 
approximate quantum encryption protocols 
and a class of schemes \cite{BHLSW03,HLSW03} 
for remote state
preparation (RSP) \cite{Lo99rsp}.  In these schemes, Alice has a
classical description of the message, applies a measurement to her
half of the ebits, and sends the outcome to Bob.  In particular, the
POVM can be chosen to contain the operators 
$\{\smfrac{1}{M}(U_k \mrho U_k^\dagger)^T\}_k$ for
$M=\| \sum_k U_k \mrho U_k^\dagger \|_\infty$, and conditioned on
receiving the outcome $k$, Bob's half of the ebits becomes $U_k \mrho
U_k^\dagger$. (See also Appendix \ref{app:transposetrick}.) 
% and \ref{sec:psqa}.) 
When $k$ takes $2^{m{+}2\log(1/\epsilon)+8}$ values, RSP succeeds with
probability at least $1-\epsilon$ \cite{HLSW03,A08}, while using only
about half of the communication required by teleportation.
Furthermore, this communication cost is optimal \cite{BHLSW03,HLSW03}.
As a result, approximate encryption requires about half of the key
needed for exact quantum encryption.  One can interpret this result as
follows.  Exact quantum encryption breaks all possible correlations
with a purifying system, while approximate encryption does not.  The
decorrelation in exact quantum encryption requires the extra key.

%-----------------------------------------------------------------------
\subsection{Quantum authentication}

\label{sec:qa}

{\bf Definition} 

Alice and Bob share a key $K$ with distribution $\{p_k\}_k$, and the
realization is $k$.  Alice applies an encoding map $\CE_k$ that takes
the $m$-qubit message system $M$ to an $(m{+}l)$-qubit system $T$,
which is transmitted to Bob.  After Bob receives the possibly altered
system $T$, he applies a decoding map $\CD_k$ which outputs an
$m$-qubit message in $M$ and one extra qubit $V$ with two states
labelled by $|\acc\>,|\rej\>$.
The security conditions apply to any purification $|\psi\>_{RM}$
(with reference system $R$) of the message state in system $M$.

\vspace*{0.5ex}

\emph{Completeness:} $\forall k$: $(\CI_R \otimes (\CD_k \, \CE_k)_M)
(\psi_{RM}) = \psi_{RM} \otimes \acc$

\emph{Soundness:} Let the adversarial attack be a TCP map given by
${\cal O}$.  Then the output of the protocol is $\tilde{\mrho}_{RMV} =
\sum_k p_k (\CI_R \otimes (\CD_k \, {\cal O} \, \CE_k)_M)
(\psi_{RM})$.  The scheme is said to have security parameter $\e$ if
$\Tr \lbm (I-\psi)_{RM} \otimes \acc_{V} \rbm \tilde{\mrho}_{RMV} <
\epsilon$.

\vspace*{1ex}

Intuitively, the above conditions say that quantum authentication
should accept and transmit a message perfectly in the absence of
tampering, and reject with high probability otherwise.  

Unlike quantum encryption, it is intrinsically impossible to 
achieve perfect soundness.  
The issue of approximate security involving purifications is subtle
but it was not explicitly dealt with in \cite{BCGST02}. (See endnote
\cite{qasecdef}.)
The above conditions take into account purifications which captures
all possible correlations between the message $M$ and other systems 
(though not obviously composable).

{\bf Known constructions} 

We first describe the quantum authentication scheme QA
constructed in \cite{BCGST02} in detail in the following.  It has two
main subroutines, the quantum encryption scheme QEnc described in the
previous subsection, and quantum purity test codes (PTC), which are 
closely related to quantum purity test protocols (PTP).

Consider a set of quantum stabilizer codes $\{Q_t\}$
\cite{Gottesman96,Gottesman97} encoding $m$ qubits into $n$ qubits.  
The set $\{Q_t\}$ is said to be a stabilizer purity test
code with error $\epsilon$ if, for any nontrivial $n$-qubit 
Pauli error $E$, at
least a fraction $1-\e$ of the codes detect it.

A purity test protocol with error $\epsilon$ is a superoperator
${\cal T}$ which can be implemented with local operations and
classical communication (LOCC), and which maps $2n$ qubits,
half held by Alice and half by Bob, to $2m+1$ qubits satisfying
the following two conditions (here $n=m+l,l\geq0$):

\emph{Completeness:} ${\cal T} (\Phi^{\ot n}) = \Phi^{\ot m} \otimes
{\acc}\, .$

\emph{Soundness:} $\forall \mrho$ $\Tr \, \lbm \! {\cal T}(\mrho)
\lbm (I-\Phi^{\ot m}) \, \ot \, {\acc} \rbm \rbm < \e$.

\vspace*{0.5ex}

Each purity test code $\{Q_t\}$ gives rise to a purity test 
protocol ${\cal T}$ as follows \cite{BCGST02}.
Each of Alice and Bob measures the syndrome of $Q_t$ on his/her $n$
qubits, for the same random $t$.  If their syndromes agree, they
accept and then perform the decoding procedure for $Q_t$; otherwise
they reject.
If the purity test code $\{Q_t\}$ has error $\epsilon$, then ${\cal
T}$ is a purity test protocol with error $\epsilon$.
An efficient purity test code was constructed in \cite{BCGST02}, 
such that $l=s$ and $\e = 2 \, \smfrac{1+m/s}{1+2^s}$ for message 
length $m$ and any chosen $s$.
The shared random variable $t$ should be independent of the $2n$-qubit
input of the purity test protocol.  
Throughout this paper, $t$ is a secret key inaccessible to the
adversary to ensure the independence condition.
In the LOCC setting it can
be generated by one party and communicated to the other party.

The noninteractive protocol QA with security parameter $\e$ consists
of first applying QEnc to the $m$-qubit message, followed by using
additional secret key to further encode with a purity test code and
then apply an operation corresponding to a random syndrome (all
parameters as described above).
Formally, QA$=$QA[QEnc,PTC,KD$_\I$].
It requires $m+s$ qubits of quantum communication and
$2m+s+\log_2(2^s+1)$ key-bits. Both costs are asymptotically optimal
as quantum encryption is necessary for quantum encryption
\cite{BCGST02}.
The security of QA is reduced to that of an interactive protocol TQA' in
which a purity test protocol is first used to establish a
$(2m{+}1)$-qubit state, the first $2m$ qubits are used to teleport the
message, and the last is in the $\acc$ or $\rej$ state.
QA satisfies the completeness and soundness security conditions stated
above. (See endnote \cite{qasecdef}.)

See Sect.~1 (under ``related works'') for some additional very recent 
constructions \cite{GYZ16}.  

%-----------------------------------------------------------------------
\subsection{Quantum Universal Composability Theorem}
\label{sec:qucompos}

Throughout the paper, we denote the associated ideal functionality of
a protocol by adding a subscript $\I$.  Different protocols can have
the same ideal functionality.  A protocol $\P$ calling a subprotocol
$\s$ is denoted as $\P[\s]$.  Two conjoining protocols (implemented by
a joint circuit) are written as $\P_1$+$\P_2$.  Two protocols $\P_1,
\P_2$ implemented with the same circuit are said to be equal $\P_1 =
\P_2$; the circuit can be interpreted in two ways.

In the {\em universal composability} (UC) approach
\cite{Canetti01,QIP03-1,BM02}:

(1) A UC security definition for a primitive is one that can be stated
for a single execution of the primitive but nonetheless guarantees
security of composition with any other properly defined system.
This definition involves a description of some ideal functionality of
the primitive.
The goal is to preserve security in a \emph{basic composition}.  
More concretely, we want a security definition such that, if $\s$ is a
secure realization of an ideal subroutine $\s_\I$, and a protocol $\P$
using $\s_\I$, written as $\P[\s_\I]$, is a secure realization of
$\P_\I$ (the ideal functionality of $\P$), then $\P[\s]$ is also a
secure realization of $\P_\I$.  

(2) A prescription for how to securely perform basic composition
recursively allows any complex protocol to be built out of secure
components.

A simplified model appropriate to our setting is described in
Appendix~\ref{app:qucmodel}. (See also \cite{BHLMO-TCC05}.)  
In essence, the UC security condition for $\P$ expresses that $\P$ and
$\P_\I$ are indistinguishable by any adversarial attack.
It does so by defining an ``environment'' $\CE$ that includes the
actual adversary and any application protocol that calls $\P$. The
environment controls the protocol's input and receives its output, and
ultimately itself outputing a binary random variable $\G$.
For this $\CE$, extend $\P_\I$ by a simulator $\S$ to an {\em extended
ideal protocol} and denote the conjoining protocols as $\P_\I{+}\S$.
$\CE$ still controls the input/output of the unit (out of the control
of $\S$) but insecure channels and other insecurities of $\P$ are
``simulated'' by $\S$.  (See Fig.~1 in Appendix~\ref{app:qucmodel}.)
The random variable $\G_\I$ output in this case generally differs from
$\G$, and their statistical difference quantifies the security -- the smaller 
the statistical difference the higher the security.
This motivates the following definition of universal composable security:

{\bf Definition 1:} $\P$ is said to $\e$-securely realize $\P_\I$
(shorthand $\P \; \e$-\sr$\P_\I$) if
\be
    \forall \CE ~~\exists \S  ~~{\rm s.t.}~~
    % \big | \,\Pr(\G{=}0) - \Pr(\G_\I{=}0) \, \big | \leq \e
    \| \, \G - \G_\I \, \|_1 \leq \e
\;.
\label{eq:usd}
\ee

We call $\e$ in \eq{usd} the {\em distinguishability-advantage}
between $\P$ and $\P_\I$.  It has a simple operational meaning.  The
entire interaction between the environment $\CE$ (including the
adversary) and the protocol $\P$ can be described by a circuit of
gates and channels, as can the interaction between $\CE$ and
$\P_I{+}\S$.  For a given environment, each interaction results in a
corresponding final state.  The environment makes the \emph{best}
quantum measurement to distinguish which one of the two final states
it has, and the output distributions for the two interactions are $\G$
and $\G_\I$ respectively.  Due to a result by Helstrom (see
\cite{Helstrom67,Helstrom76}) the maximum value of $\| \, \G - \G_\I
\, \|_1$ is simply the trace distance between the two possible final
states (before the measurement).  Thus $\e$ is an upper bound to the
trace distance between the possible final states, maximized by the
environment and minimized by the simulator.

This security definition (in the model described) is useful because
security of basic composition follows ``by definition''
\cite{QIP03-1,BM02}.
\begin{theorem} Suppose a protocol $\P$ calls a subroutine $\s$.
If $\s$ $\e_\s$-\sr$\s_\I$ and $\P[\s_\I]$ $\e_\P$-\sr$\P_\I$, then
$\P[\s]$ $\e$-\sr$\P_\I$ for $\e \leq \e_\P{+}\e_\s$.
\label{thm:basic-com}
\end{theorem}
Theorem 1 can be generalized to any arbitrary protocol with a proper
{\em modular structure}, as defined in Appendix \ref{app:qucmodel}.
An example of an improper modular structure is one with a security
deadlock, but the protocols we analyze in this paper all generate
proper modular structures.
The idea is to represent the protocol as a tree and then apply
Theorem \ref{thm:basic-com} recursively to the leaves of the tree.
Roughly speaking, to build the tree, represent any arbitrary
protocol $\P$ using subprotocols $\{\s_{i}\}$ by a $1$-level tree,
with $\P$ being the parent and $\{\s_{i}\}$ the children.
Recursively replace these children by trees until the leaves are the
basic primitives subject to analysis, and call this the associated
tree of $\P$.
(More general modular structures,
represented by acyclic directed graphs, can be transformed into
trees~\cite{BM02}.)
Then the security of $\P$ can be stated in terms of that of the
components in the tree:
\begin{theorem} Let $\P$ be a protocol and $T_\P$ its associated
tree.
For each vertex $v$ in $T_\P$, let ${\cal M}_v$ be the subprotocol
corresponding to $v$ with its own subprotocols 
$\{{\cal N}_i\}_{i=1,\cdots,l}$.  (This can be an empty set if $v$ is
a leave.)
Then, if ${\cal M}_v[{\cal N}_{1\I},\cdots,{\cal N}_{l\I}]$
$\e_{{\cal M}_v}$-\sr${\cal M}_\I$, we have $\P$ $\e$-\sr$\P_\I$ for $\e
\leq \sum_{v} \e_{{\cal M}_v}$.
\end{theorem}
Theorem 2 is obtained by the recursive use of Theorem 1 and the
triangle inequality, replacing each subprotocol by its ideal
functionality, from the highest to the lowest level (from the 
leaves toward the root).
The distinguishability-advantage between $\P$ and $\P_\I$ is upper
bounded by the sum of all the individual distinguishability-advantages
for the replacements.

It is worth mentioning that there is an alternative to Definition 1
above for the universal composable security definition: 

{\bf Definition 2:} $\P$ is said to $\e$-securely realize $\P_\I$
(shorthand $\P \; \e$-\sr$\P_\I$) if
\be
    \exists \S ~~{\rm s.t.} ~~ \forall \CE ~~
    % \big | \,\Pr(\G{=}0) - \Pr(\G_\I{=}0) \, \big | \leq \e
    \| \, \G - \G_\I \, \|_1 \leq \e
\;.
\label{eq:usd3}
\ee
In other words, the order of the quantifiers has been exchanged in
this alternative.  Definition 2 offers stronger security than
Definition 1.  The basic composition law holds for each definition --
UC-secure primitives satisfying Definition 1 give composition with like
security, and similarly for Definition 2.  However, when composing
protocols with mixed security definitions, the composition generally
satisfies the weaker definition.

Definition 1 is used in the Ben-Or--Mayers model.  Their analysis
still holds for definition 2 for composition involving a constant
number of components.

In our work, we prove UC-security for EBIT[PTP], TQA[EBIT[PTP],C$_\I$]
and QA+KG according to definition 2.  Thus simple applications of
these protocol will inherit the stronger security.

%%%%%%%%%%%%%%%%%%%%%%%%%%%%%%%%%%%%%%%%%%%%%%%%%%%%%%%%%%%%%%%%%%%%%%%%%
\section{Universal Composable Security for QA+KG}
\label{sec:proof}

We now show the UC-security of QA+KG in the universal composability
framework.  As discussed after Definition 1, we can describe the
interaction between the environment $\CE$ and the protocol $\P$ (or
$\P_\I {+}\S$) by a circuit of gates and channels.  The
distinguishability advantage is just the trace distance between the
possible final states, maximized by the environment and minimized by
the simulator.  The circuit representation of the interaction is a
very concise summary of the state at each stage of the interaction.
Moreover, if we replace one circuit component by another, we can
capture the difference induced on the state right after that
component, and additional circuit elements cannot increase the trace
distance (by the monotonicity of the trace distance under quantum
operations).

With the above in mind, our proof consists of the following steps: \\[0.5ex]
(1) Show that for any environment, the interactions with QA+KG and
with TQA+KG result in the same final state, and they are therefore
completely indistinguishable to the environment.  In other words,
QA+KG $0$-s.r. TQA+KG.  \\ Recall that QA uses QEnc and PTC as
subroutines, as well as secret keys (given by ideal key distribution
boxes KD$_\I$).  In QA+KG, the encryption key is recycled if 
QA accepts the message.  So we may
express the first protocol QA+KG as (QA+KG)[QEnc,PTC,KD$_\I$].  The
second protocol TQA+KG first creates entanglement using PTC via the
insecure channel (but Bob never tells Alice if the entanglement is
accepted or rejected), next teleports the quantum message from Alice
to Bob (using an ideal classical channel which is authenticated,
encrypted, and hidden) and finally outputs the Bell measurement
outcome in teleportation as a new key, if PTC accepts. 
Thus, TQA+KG can be expressed
as (TP+KG)[EBIT[PTC],C$_\I$]. \\[0.5ex]
(2) Re-write the circuit for EBIT[PTC] as a circuit for EBIT[PTP], a
protocol creating ebits using a purity test protocol, such that the 
two circuits are completely indistinguishable to the environment.  
Therefore, EBIT[PTC] $0$-s.r. EBIT[PTP].  
\\[0.5ex]
(3) Show that EBIT[PTP] $(2 \sqrt{2} \e^{1/3})$-s.r. EBIT$_\I$ using the
soundness condition of EBIT[PTP].  Here, $\e \leq 2^{-s'}$ is the upper
bound of probability of failure in PTP. \\[0.5ex]
(4) By (2) and (3), applying Theorem \ref{thm:basic-com} with $\P$=TP,
$\s$=EBIT[PTC], it follows that TP[EBIT[PTC],C$_\I$]
$(2 \sqrt{2} \e^{1/3})$-s.r. TP[EBIT$_\I$,C$_\I$] which $0$-s.r. Q$_\I$,
the ideal functionality of the perfectly authenticated and encrypted
quantum channel.  \\[0.5ex]
(5) Show that (TP+KG)[EBIT[PTC],C$_\I$] $2 \sqrt{2} \e^{1/3}$-s.r.
(TP+KG)[EBIT$_\I$,C$_\I$] $0$-s.r. Q$_\I$+KD$_\I$, where KD$_\I$ is
the ideal functionality for generating a key of a certain size between
Alice and Bob.

{\bf Overall result:}\\ Putting (1), (5), and (4) together, 
(QA+KG)[QEnc,PTC,KD$_\I$]
$0$-s.r. (TP+KG)[EBIT[PTC],C$_\I$]  
$2 \sqrt{2} \e^{1/3}$-s.r. (TP+KG)[EBIT$_\I$,C$_\I$] 
$0$-s.r. Q$_\I$+KD$_\I$.  

So, (QA+KG)[QEnc,PTC,KD$_\I$] $2 \sqrt{2} \e^{1/3}$-s.r. Q$_\I$+KD$_\I$.

\vspace*{1ex}

\clearpage

We now prove these steps.  Consider Circuit Diagram 1 below, with 
a schematic diagram for QA+KG and its interaction with the
environment:\\[0ex]
\be
\epsfig{file=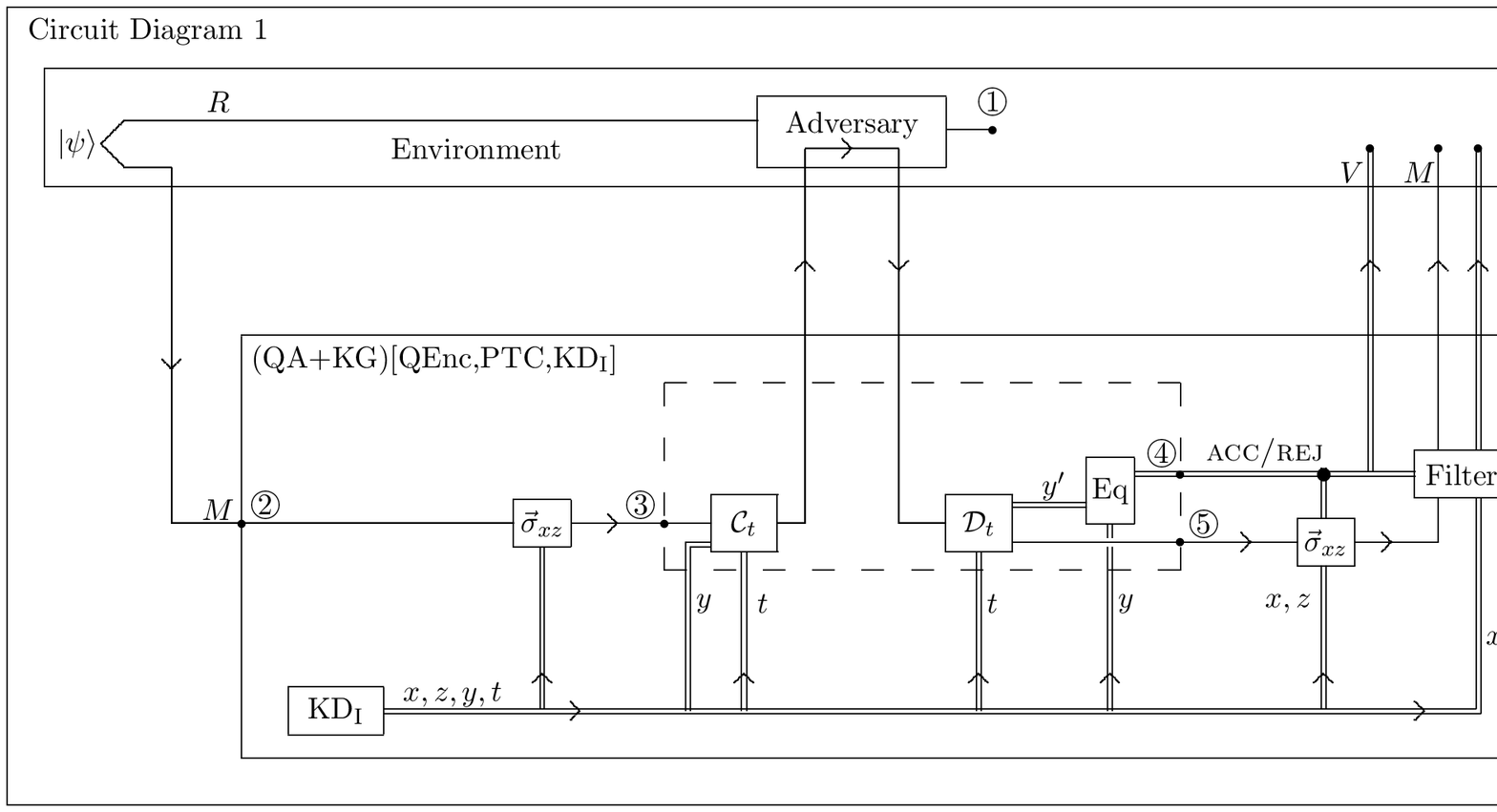,width=6.2in}
\nonumber
\ee

Our {\em circuit diagrams} use the following conventions.
(See \cite{Nielsen00bk} for more detail.)  Time runs from left to right.  
The box around the
environment denotes what is accessible to it.  Single and double lines
represent quantum and classical information (moving in time or space)
respectively.  Additional arrows explicitly indicate the direction of
information flow.  Small boxes with input/output information denote
operations.  Such diagrams are concise descriptions of the protocols
and summarizes how the states evolve.  They will be used to present a
significant part of our proofs.
To help readers gain familiarity with this representation, we go through
the above diagram in detail.

We consider the most general environment allowed in the universal
composability model.  The environment chooses an arbitrary state
$|\psi\>_{RM}$.
The system $M$ carries the quantum message to be authenticated, 
and $R$ carries all possible correlations to the quantum message.  
The environment supplies the register $M$ as input to the analyzed
protocol (in particular, as input to QA), by communicating $M$ to
Alice.   See the far left of circuit diagram 1.  
In the security definition for quantum authentication in Section
\ref{sec:qa}, the system $R$ is left unchanged and is used as a
reference for checking that the correlations to the message $M$ are
preserved.
However, when considering universal composability, the environment can
share data with the adversary.  Therefore, in our analysis, the system
$R$ is given to the adversary who can use it as quantum side
information when attacking the transmission in QA.
After the attack, the adversary passes all data back to the
environment (labeled \textcircled{1}).

We now turn to the analyzed protocol QA+KG.  
We model the perfect keys used by the protocol by including an ideal
key distribution box labeled KD$_\I$, which distributes a perfect key
(with $4$ parts, $x,z,t,y$) between Alice and Bob that is not
accessible to the environment.
As a side remark, note that we can model components of a protocol
mathematically with perfect devices that need not be realized
physically.  It simplifies further analysis when the perfect key comes
from an actual protocol; we only need to check the universal
composable security of the latter \cite{BHLMO-TCC05,RK05} (and apply
Theorem 1).

As described in Section \ref{sec:qa}, Alice encrypts the message in
system $M$ with $\vec{\s}_{xz}$ (the $m$-qubit Pauli operator
specified by the two $m$-bit keys $x$ and $z$) and then applies the
purity test code by encoding in an error correcting code indexed by
key $t$ and injecting an error syndrome indexed by key $y$.  Without
loss of generality, this encoding operation is a unitary
transformation ${\cal C}_t$ that acts jointly on the input logical
state and the syndrome $y$.  The encrypted and encoded state 
is then transmitted to the adversary (part of the environment here),
which can attack it jointly with $R$ using any physical process.
After that, the possibly altered state is received by Bob.  He first
applies ${\cal D}_t$, which reverses ${\cal C}_t$ to return some
quantum state \textcircled{5} and a syndrome $y'$.  Then $y$ and $y'$
are compared in the ``Eq'' operation, which outputs in register $V$
the state {\acc} if $y=y'$ and {\rej} otherwise.  Bob also decrypts
the quantum output of ${\cal D}_t$ (\textcircled{5}) with $\vec
\s_{xz}$, producing a final quantum message in system $M$.
Finally, the {\rej} state will trigger the filter to replace the final
quantum message and the keys $x,z$ by error symbols, else the systems
$VM$ will be the output for QA, and the keys $x,z$ will be the output
for KG.

{\bf Step (1):~} Showing QA+KG $0$-s.r. TQA+KG=(TP+KG)[EBIT[PTC],C$_\I$]. 

Consider Circuit Diagram 2 below, with a schematic diagram for 
TQA+KG and its interaction with the environment.\\
\be
\epsfig{file=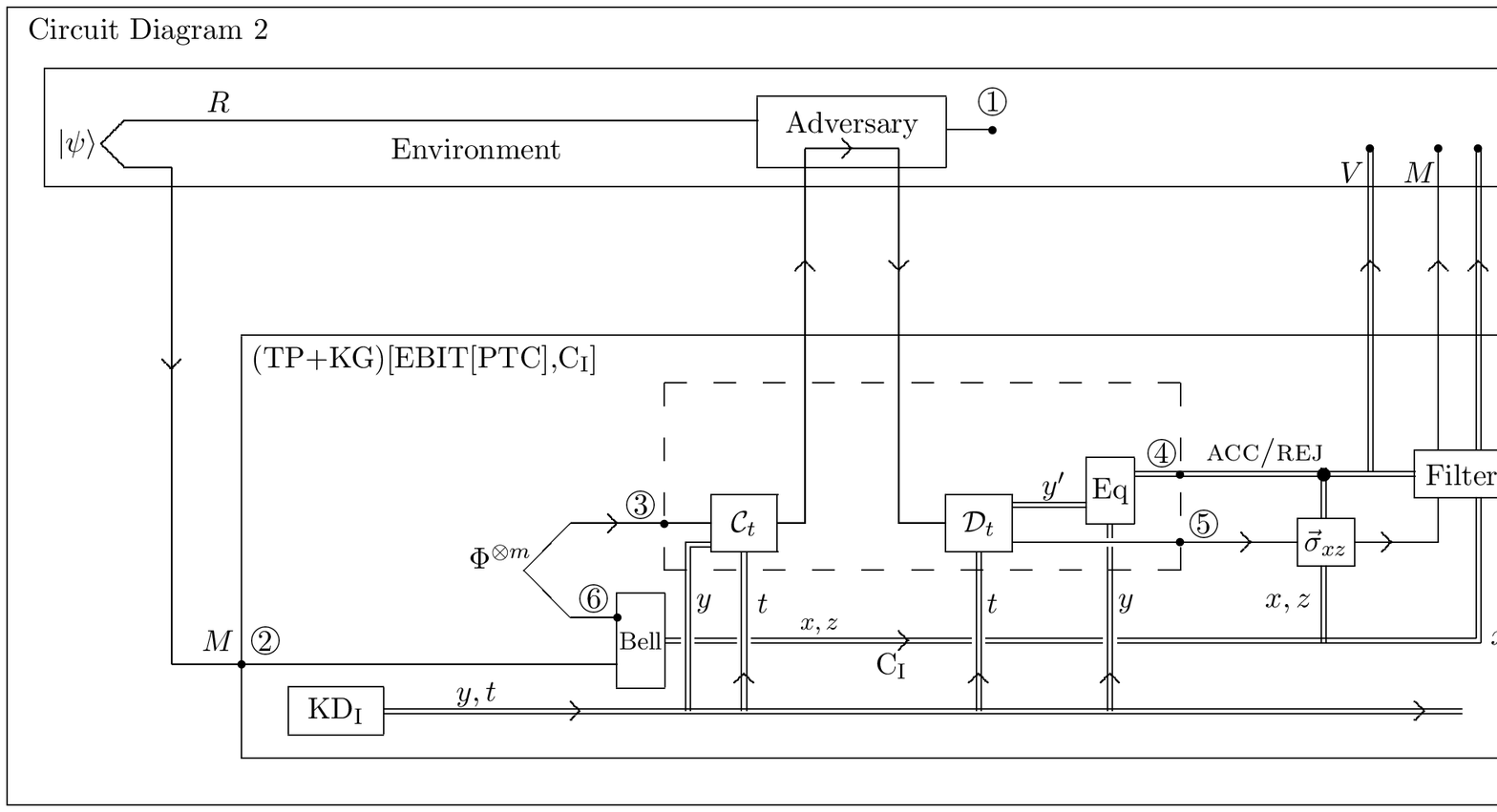,width=6.2in}
\nonumber
\ee

Circuit diagrams 1 and 2 (for the interaction between the environment
and QA+KG and TQA+KG) differ in only two places.  First, encryption in
QA+KG is done in TQA+KG by Alice preparing $m$ ebits $\Phi^{\ot m}$
(all $2m$ qubits with her) and applying Bell measurements on the $m$
halves of the ebits and the incoming message $M$.  Having measurement
outcomes $x,z$ in TQA+KG is the same as applying $\vec \s_{xz}$
directly to the message in QA+KG.
The states labeled \textcircled{3} are identical in both protocols, 
though prepared differently.  More specifically, $(x,z)$ arises  
differently, but it is 
completely random in both protocols, and for a given $(x,z)$, the
postmeasurement states in \textcircled{3} are identical.  
Second, the measurement outcomes $x,z$ are {\em communicated} to Bob by a
{\em hidden} channel C$_\I$ whose execution and content are unknown by
anyone except for Alice and Bob, and the content is transmitted
exactly.
Such an unrealistic resource can be part of an ideal or partially
ideal functionality against which we are comparing the actual
protocol.  Neither change will affect the final state of the
environment, and thus the two interactions are identical from the
point of view of the environment.

Now, in TQA+KG, Alice's Bell measurement can {\em formally} be delayed
until after the ``Eq'' operation. Thus, the dashed box in Circuit Diagram 
2 with input
\textcircled{3} and outputs \textcircled{4}, \textcircled{5} is only
used in transmitting half of $\Phi^{\ot m}$ (the other half being
\textcircled{6}).  
So, TQA+KG can be interpreted as teleportation TP calling EBIT[PTC] in
addition to the hidden channel C$_\I$ which is perfectly encrypted and
authenticated classical channel as subroutines.  We write TQA+KG =
(TP+KG)[EBIT[PTC],C$_\I$], to emphasize the modular structure and the
potentially insecure components.  Note that in this version of
EBIT[PTC], Alice does not know whether Bob accepts or rejects, since 
we want to make a correspondence with the {\em noninteractive} QA.  

{\bf Step (2):~} Re-expressing the circuit for EBIT[PTC] as a circuit
for EBIT[PTP].  

Consider EBIT[PTC], the components in TQA+KG that start with 
the state labeled by \textcircled{3}\textcircled{6} and 
led to the state
labeled by
\textcircled{4}\textcircled{5}\textcircled{6}.
There is no input message, and it simply creates $m$ ebits using the
purity test code.
We extract it as the left diagram in the following (Circuit Diagram E1) 
for the analysis of its UC security.
We include in EBIT[PTC] the register $V$ that holds the measurement
result \acc~or \rej, which is known to Bob.  Alice will not know if
Bob accepts or rejects in our application, but our analysis also holds
for interactive protocols.  The filter operation replaces 
\textcircled{5} by an error symbol if $V$ is in the state \rej.  
(For interactive protocols, \textcircled{6} will also be replaced by
an error symbol.)  
We use $A$ and $B$ to denote the registers holding the final output 
``EPR pairs'' (systems \textcircled{6} and \textcircled{5} in EBIT[PTC]).  
The circuit components of EBIT[PTP] (in Circuit Diagram E2) are
defined similarly, except $y$ is now a measurement outcome that is
communicated from Alice to Bob using a perfect classical channel.
The subroutine EBIT[PTP] is only used in the analysis and the 
requirement to send $y$ does not play a big role, 
so, we omit the explicit subroutine label C$_\I$ for simplicity.

\be
\epsfig{file=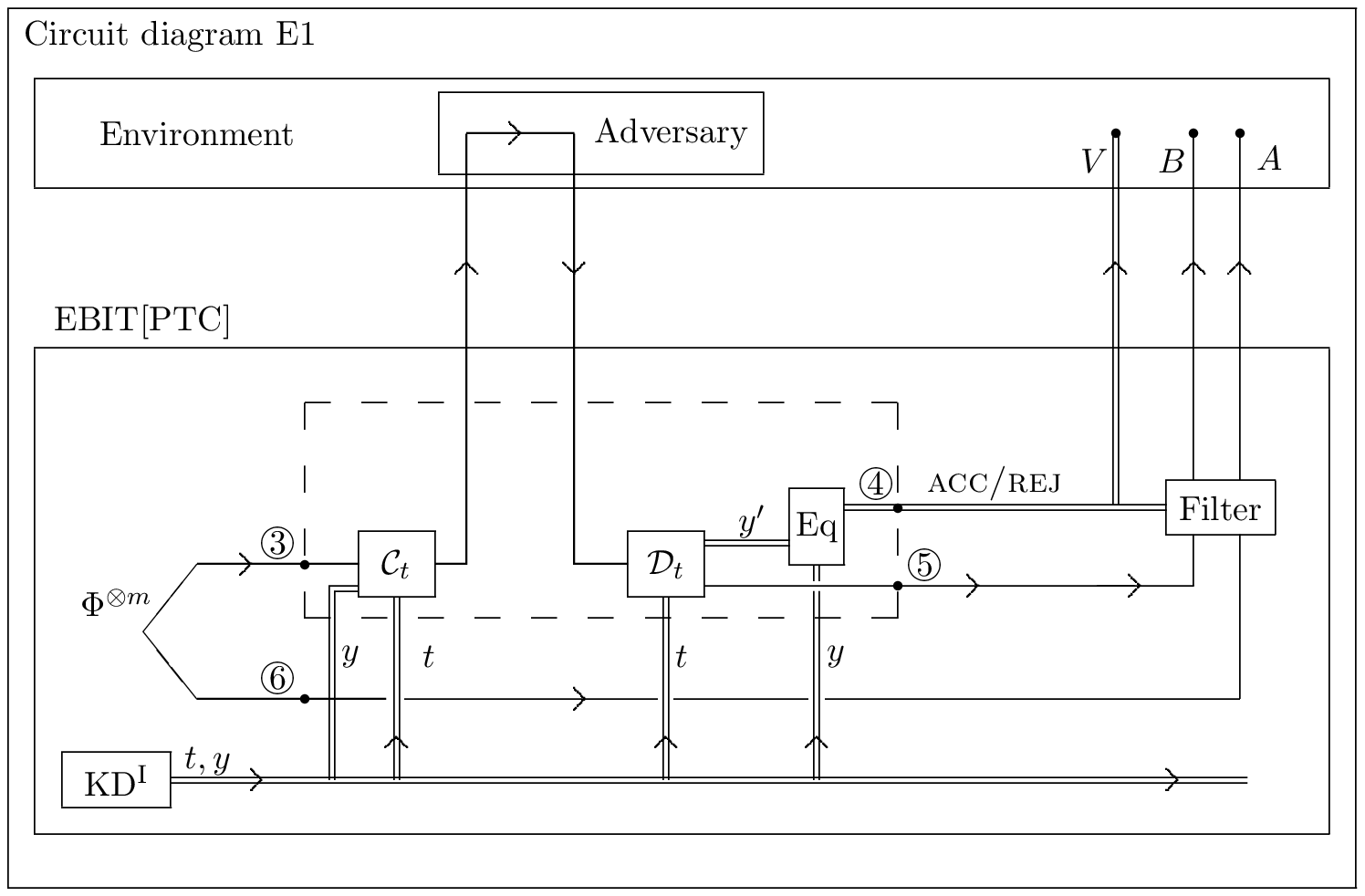,width=3.3in}
~\epsfig{file=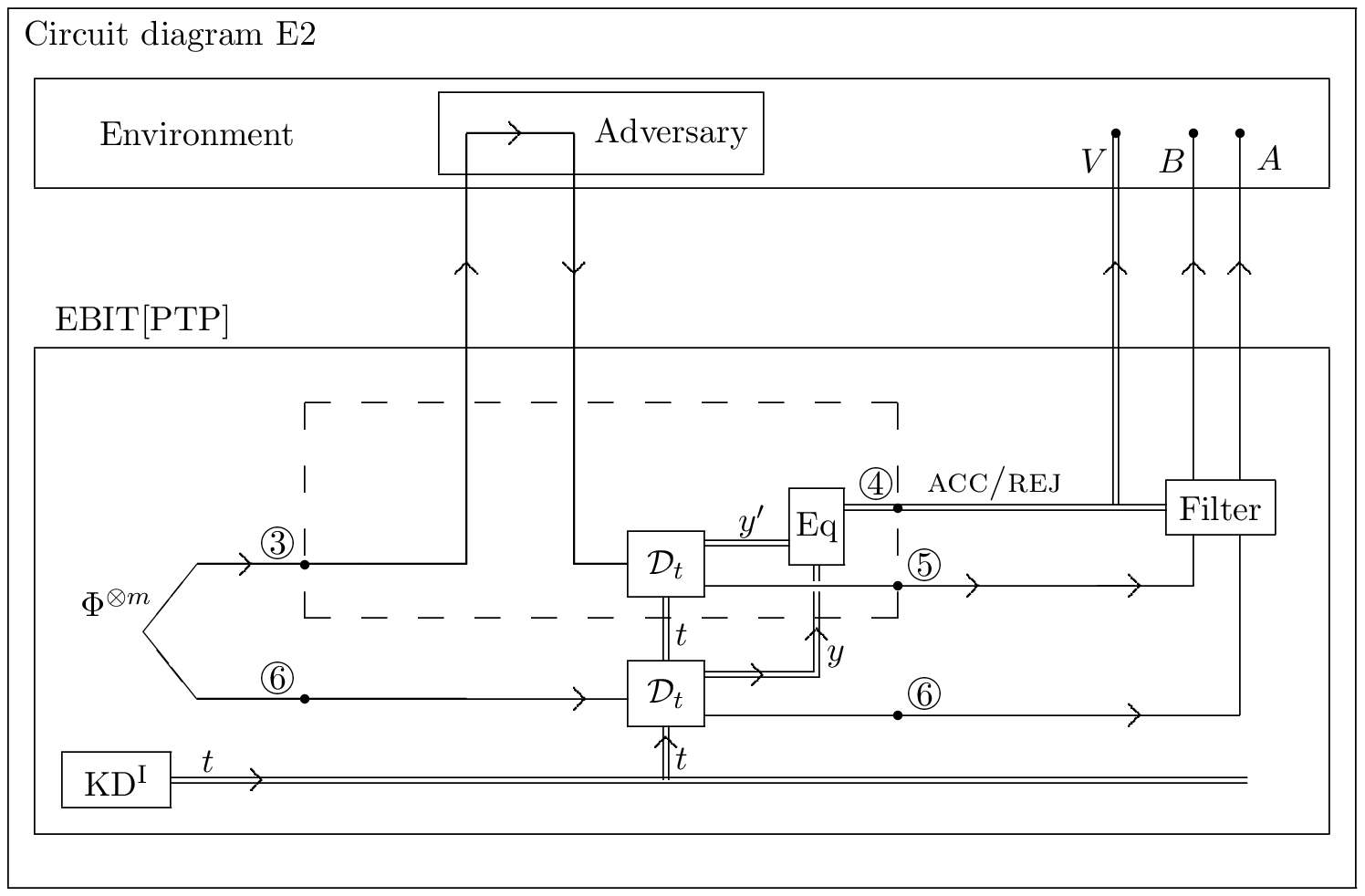,width=3.3in}
\nonumber
\ee
Now, the states labeled by
\textcircled{4}\textcircled{5}\textcircled{6} are the same in
EBIT[PTC] (circuit diagram E1, left)  
and EBIT[PTP] (circuit diagram E2, right) 
shown above.  This was proved in Appendix E of
\cite{BCGST02} and we provide an elementary proof in Appendix
\ref{app:transposetrick}.  

\clearpage

{\bf Step (3):~} Showing EBIT[PTP] $(2 \sqrt{2} \e^{1/3})$-s.r. EBIT$_\I$.

\be
\epsfig{file=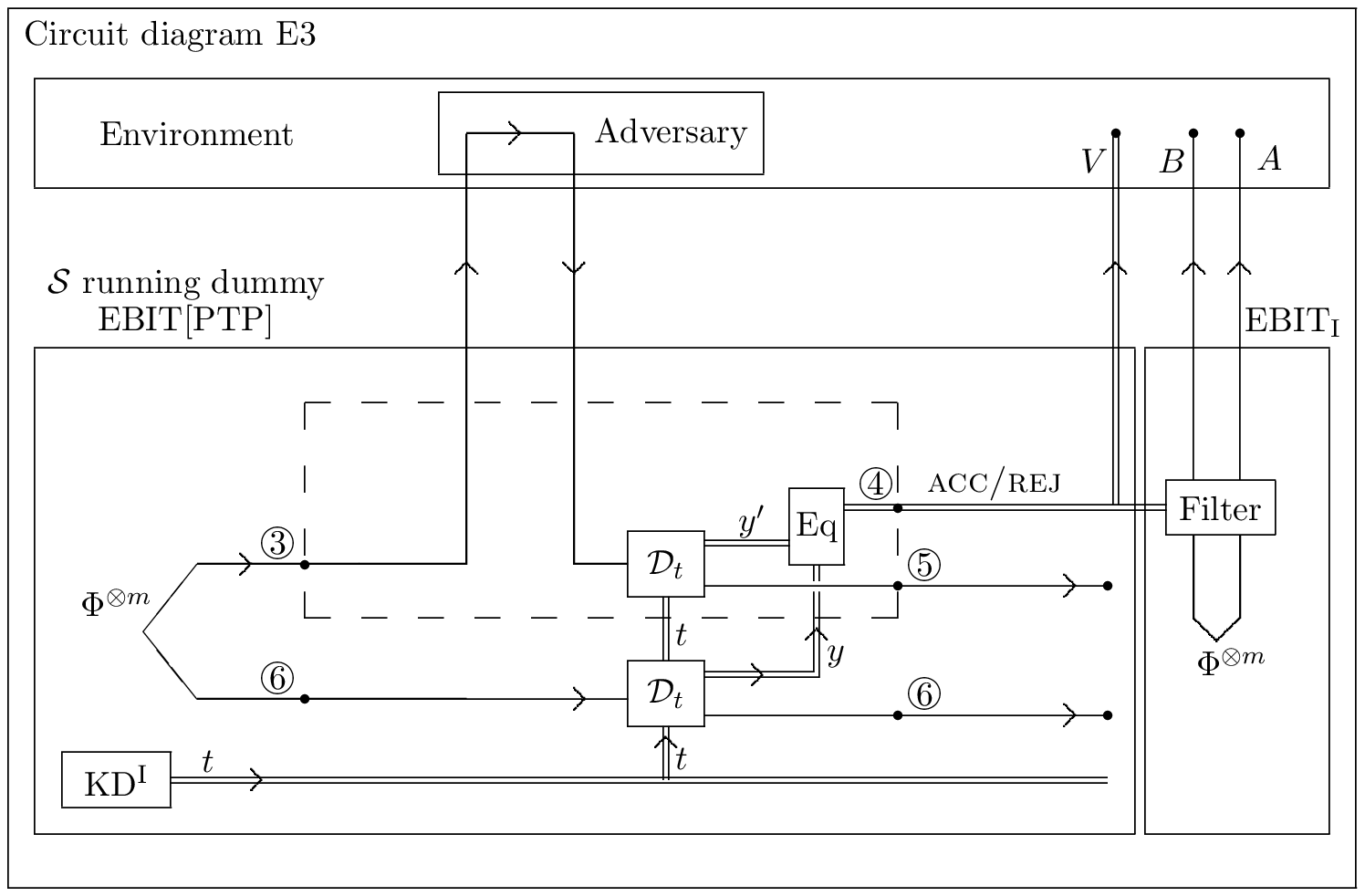,width=3.3in}~~
\raisebox{1.0in}{\parbox{3.0in}{
We now analyze EBIT[PTP] in the UC
framework against the ideal protocol EBIT$_\I$ defined as follows. 
EBIT$_\I$ takes an input in the state
{\acc} or {\rej} and outputs $m$ perfect ebits or an error
state {\sc err} accordingly.  
In this paper, the state {\sc err} is a tensor product of a maximally
mixed state on system $A$ and an error symbol on system $B$.  (For 
interactive protocols, both $AB$ can be in error symbols.)  This 
choice for {\sc err} will minimize the distinguishability advantage 
between EBIT[PTP] and EBIT$_\I$.}}
\nonumber
\ee

The simulator $\S$ runs a ``dummy'' execution
of EBIT[PTP] and takes the dummy {\acc} or {\rej} state and feeds it
into EBIT$_\I$.  Note that this $\S$ is independent of
the environment.  So, the quantifiers are as given by \eq{usd2}
in Definition 2.%}}

The distinguishability advantage can be upper bounded by the trace
distance between the two states held by the environment after the
executions of EBIT[PTP] and $\S$+EBIT$_\I$.  Let these states
be denoted by $\eta^{\mbox{\sc pt}}$ and $\eta^{\I}$.  Then we can write
\\[0.2ex]
\bea
\begin{array}{rccccccccl}
\eta^{\mbox{\sc pt}} & = &
   p_{\acc} & \xi_{ABE} & \ot \; \acc &
 + & p_{\rej} & \mbox{\sc e \hspace*{-1.8ex} r \hspace*{-1.85ex} r}_{AB} 
   & \ot \mu_E & \ot \rej
\\
\eta^\I & = &
   p_{\acc} & \Phi^{\ot m} \ot \xi_{E} & \ot \; \acc &
 + & p_{\rej} & \mbox{\sc e \hspace*{-1.8ex} r \hspace*{-1.85ex} r}_{AB} 
   & \ot \mu_E & \ot \rej
\end{array}
\label{eq:etas}
\eea
\vskip 1ex
where $\xi_E = \Tr_{\!AB} \, \xi_{ABE}$ (likewise, $\xi_{AB} =
\Tr_{\!E} \, \xi_{ABE}$, and similarly for $\eta$), and {\sc err}$_{AB}$ 
is the tensor product of a maximally mixed state on $A$ and an error
symbol on $B$, chosen so that the \rej~ terms are identical in 
$\eta^{\mbox{\sc pt}}$ and $\eta^\I$.
Note that $p_{\acc}$, $p_{\rej}$, $\xi_{ABE}$, and $\mu_E$ are determined 
both by the adversarial attack and the PTP.  Our choice of the simulator 
ensures that the reduced states on $E$, conditioned on each of $\acc$ and 
$\rej$, are the same for $\eta^{\mbox{\sc pt}}$ and $\eta^\I$.

{From} \eq{etas}, $\| \, \eta^{\mbox{\sc pt}} \, {-} \, \eta^\I \, \|_1 
= p_{\acc} \, \| \, \xi_{ABE} - \Phi^{\ot m}_{AB} \, \ot \, \xi_E \, \|_1$.
We upper bound this for $p_{\acc} \leq \epsilon^{1/3}$ and
$p_{\acc} > \epsilon^{1/3}$ separately.
If $p_{\acc} \leq \epsilon^{1/3}$, then, 
the bound is $2 p_{\acc} \leq 2 \epsilon^{1/3}$
since the trace distance is at most $2$.
If $p_{\acc} > \epsilon^{1/3}$, we seek an upper bound for   
$\|\, \xi_{ABE} - \Phi^{\ot m}_{AB} \, \ot \, \xi_E \, \|_1$. 
This is equivalent to finding a lower bound for 
the fidelity $F(\xi_{ABE}, \Phi^{\ot m}_{AB} \ot \xi_E)$. 
Since the fidelity is the maximum overlap squared between all purifications,
any specific purifications of $\xi_{ABE}$ and $\xi_E$ give a lower
bound.
Any purification of $\xi_{ABE}$ can be expressed as \\[0.1ex]
\be 
|\xi\>_{ABER} = \sqrt{1-\alpha} \; |\Phi\>^{\ot m}_{AB}
\ot |a\>_{ER}\, {+} \, \sum_{i} \sqrt{\alpha_i} \;  |\Psi_i\>_{AB} \ot |b_i\>_{ER}
\label{eq:pur}
\ee 
where $|\Phi\>^{\ot m}$ and $\{|\Psi_i\>\}_i$ form a basis for
$(\C^{2})^{\otimes 2m}$, $|a\>$ and $|b_i\>$ are unit vectors, and 
$\alpha = \sum_i \alpha_i$.
Furthermore, $\xi_{ER} = (1-\alpha) |a\>\<a| + \sum_i \alpha_i
  |b_i\>\<b_i|$, so, $F(|a\>\<a|,\xi_{ER}) = \<a| \xi_{ER} |a\> \geq
  1-\alpha$.  Now, $F(\Tr_R \,|a\>\<a|,\xi_{E}) \geq
  F(|a\>\<a|,\xi_{ER}) \geq 1-\alpha$, so, there exists a purification
  $|a^*\>$ of $\xi_{E}$ such that $|\<a^*|a\>|^2 \geq 1-\alpha$. 
Using this and \eq{pur}, we obtain 
$| \left( \<\xi|_{ABER} \right) 
   \left( |\Phi^{\ot m}\>_{AB} \ot |a^*\>_{ER} \right) |^2 \geq (1-\alpha)^2$
so,
% the overlap square of $|\xi\>_{ABER}$ in \eq{pur} and 
% $\Phi^{\ot m}_{AB} \ot |a^*\>_{ER}$ is $(1-\alpha)$\DL{$^2$} so,
$F(\xi_{ABE}, \Phi^{\ot m}_{AB} \ot \xi_E) \geq (1-\alpha)^2$.
We now show that $\alpha \leq \e^{2/3}$.
To do so, note that the soundness condition of the purity test protocol 
in this context can be expressed as   
$\Tr \lbm (\Tr_E \, \eta^{\mbox{\sc pt}}) \; 
((I{-}\Phi^{\ot m}) \ot \acc) \rbm \leq \e$.
A direct substitution of \eq{etas} into the condition gives \\[-1ex]
\be
\Tr \lbm \xi_{AB} (I{-}\Phi^{\ot m}) \rbm \leq \e/p_{\acc} < \e^{2/3} \,.
\label{eq:star}
\ee
\\[-1ex]
We can obtain $\xi_{AB}$ from \eq{pur} by tracing out $ER$ in
$|\xi\>_{ABER}$; substituting this $\xi_{AB}$ into \eq{star} 
gives $\alpha \leq \e^{2/3}$, as claimed.  
Together, 
\bea
\|\xi_{ABE} - \Phi^{\ot m}_{AB} \ot \xi_E \|_1
\stackrel{\mbox{\cite{FG97}}}{\leq}
% \sqrt{1-F(\xi_{ABE}, \Phi^{\ot m}_{AB} \ot \xi_E)^2} \leq 
2 \sqrt{1-F(\xi_{ABE}, \Phi^{\ot m}_{AB} \ot \xi_E)}
\leq 2 \sqrt{2\alpha} \leq 2 \sqrt{2} \e^{1/3}
\nonumber
\eea
\vskip 0.5ex
Thus, EBIT[PTP] $(2 \sqrt{2} \e^{1/3})$-s.r. EBIT$_\I$.

{\bf Step (4):~} By theorem \ref{thm:basic-com} and the above, 
we conclude that TP[EBIT[PTC],C$_\I$]
$(2 \sqrt{2} \e^{1/3})$-s.r.  TP[EBIT$_\I$,C$_\I$]. 
Note that TP[EBIT$_\I$,C$_\I$] in turn can serve as our
definition of the ideal functionality Q$_\I$.  The ideal
functionality Q$_\I$ uses EBIT$_\I$ and C$_\I$ as subroutines, takes input $M$
(the quantum message) and has two outputs $V$ and $M$.  If the
subroutine EBIT$_\I$ outputs \rej, Q$_\I$ also outputs \rej \, in $V$ and an
error message in $M$.  If EBIT$_\I$ outputs \acc, then, Q$_\I$ provides 
an encrypted, authenticated, and hidden channel that transmits $M$.

{\bf Step (5):~} Showing (TP+KG)[EBIT[PTC],C$_\I$] $2 \sqrt{2}
\e^{1/3}$-s.r. (TP+KG)[EBIT$_\I$,C$_\I$] $0$-s.r. Q$_\I$+KD$_\I$.
% \\ % 

Consider Circuit Diagram 2 again, and   
replace the subroutine EBIT[PTC] by EBIT$_\I$+$\S$
so the analyzed protocol becomes (TP+KG)[EBIT$_\I$,C$_\I$]. 
By step (3), (TP+KG)[EBIT[PTC],C$_\I$] $2 \sqrt{2}
\e^{1/3}$-s.r. (TP+KG)[EBIT$_\I$,C$_\I$].
We now show that (TP+KG)[EBIT$_\I$,C$_\I$] $0$-s.r. Q$_\I$+KD$_\I$.
In Q$_\I$+KD$_\I$, if Bob outputs \rej, Alice will hold a uniformly
random variable in her key output system while Bob will hold an error
message in his key output system and also in $M$.  (If Bob can send a
bit to Alice, she will replace her random variable by an error message
as well.)

First, the \acc~event occurs with the same probabilities in
both (TP+KG)[EBIT$_\I$,C$_\I$] and Q$_\I$+KD$_\I$, and similarly for 
the \rej~event.  When Bob outputs
\rej, for both protocols, we have the reduced state in
\textcircled{1}, tensor product with an error message in $M$, a
uniformly random output on Alice's key system, and an error message on
Bob's key system. When Bob outputs \acc, $M$ is teleported perfectly
in (TP+KG)[EBIT$_\I$,C$_\I$]. Furthermore, the measurement outcome
$(x,z)$ is uncorrelated with $RM$ after the operation $\vec{\s}_{x,z}$
and uncorrelated with everything else the environment has, so it is
indistinguishable from an ideal key.  Thus (TP+KG)[EBIT$_\I$,C$_\I$]
$0$-s.r. Q$_\I$+KD$_\I$.

\vspace*{1ex}

{\bf Overall result:}\\ Putting (1), (5), and (4) together, 
(QA+KG)[QEnc,PTC,KD$_\I$]
$0$-s.r. (TP+KG)[EBIT[PTC],C$_\I$]  
$2 \sqrt{2} \e^{1/3}$-s.r. (TP+KG)[EBIT$_\I$,C$_\I$] 
$0$-s.r. Q$_\I$+KD$_\I$.  

So, (QA+KG)[QEnc,PTC,KD$_\I$] $2 \sqrt{2} \e^{1/3}$-s.r. Q$_\I$+KD$_\I$.

%%%%%%%%%%%%%%%%%%%%%%%%%%%%%%%%%%%%%%%%%%%%%%%%%%%%%%%%%%%%%%%%%%%%%%%
\section{Discussion and open problems}
\label{sec:discussion}

{\bf Other results:} See appendices \ref{sec:psqa} and \ref{sec:wc}
for proofs.

First, a variant of QA+KG called PSQA+KG is UC secure when the message
is known to be pure, such as when Alice prepares it herself.  PSQA uses
the approximate encryption scheme $\approx$QEnc in place of QEnc, and
$\approx$QEnc uses only half the key needed in QEnc.
The proof involves a variant of TQA+KG in which a remote state
preparation (RSP) scheme is used in place of TP.

The Wegman-Carter classical authentication scheme is UC secure in the
quantum composability framework.  This is important in the light of
the frequent need of authenticated classical channels in many quantum
cryptographic protocols.  We provide a short informal proof, and refer 
to \cite{P13} for a more complete discussion.  

\vspace*{2ex} 

{\bf Discussion of our results and further open questions}

Using $1$ bit of back communication, and in the absence of a detected
attack, the key costs of authentication can be made negligible.  We
have also discussed two other methods to reduce the key cost in
Sect.~1.  In terms of communication cost, both QA+KG and TQA+KG are
superior to the QKD-based methods discussed in Section
\ref{sec:motivation}.
The main drawback of both QA+KG and TQA+KG is that they require a
large initial key.  This can be circumvented at the expense of a
slowly growing round complexity, however -- one can divide the
$m$-qubit messages into $\sqrt{m}$ groups of $\sqrt{m}$ qubits each
and apply QA+KG to each {\em sequentially}.
Besides the $m$ qubits of quantum communication, all other resources,
including the initial key and all classical communication, will be
sublinear in $m$.
Similarly, TQA+KG can be used instead, but an extra $2m$ classical
bits of forward communication is needed to gain the extra data
protection.

A remaining question is the extent to which recycling can be done in
QA+KG when the authentication output is {\rej}.  Ref. \cite{BRICS05},
which considers encryption of classical messages, proved that at least
$m{-}1$ key-bits have to be discarded.
Because of superdense coding, our intuition is that one may need to
discard roughly $2m$ key-bits for unknown quantum messages.  
Surprisingly, partial key recycling is recently proved to be possible
in \cite{P16}.

At the other extreme, \cite{BRICS05} found that the entire key, even
the authentication tag, can be recycled in their scheme (by using much
more key to start with) in the case of \acc.  More recently,
\cite{GYZ16} proposed quantum authentication scheme with the same
feature (though their schemes also require more key than QA), and
\cite{P16} proved that the entire key can be recycled in QA.  While
qualitatively interesting,
this question is not of practical importance; we could easily make up
for this small extra recycled key without it by having Alice append
that number of ebit-halves to the message and keeping the other halves
herself.  Upon passing the authentication test, this will make up for
the consumed keys $t,y$ and more if so desired.  Meanwhile, the number
of ebits is sublinear in the message size so the resource counting is
unaffected.

In fact, Bennett posed to us the question of using QA+KG as a simple
means to perform QKD.  QKD can resist very high transmission noise and
eavesdropping.  The main challenge here is that noise is not dealt
with efficiently by PTC, which is an error detecting code rather than
an error correcting code.  The solution turns out very simple and
there are two equivalent ways to see it.  

More generally, consider the case of quantum message authentication
through uses of a noisy channel ${\cal N}$.  This information can be
given to Alice and Bob, or they can make reasonable estimates and
assumptions about the underlying channel.  The adversary can further
tamper with the transmission.
Alice and Bob should agree on an error correcting code.  
Let ${\cal N}'$ denote the composition of the encoding, transmission
by ${\cal N}$, and decoding.  We require ${\cal N}'$ to approximate
$(m{+}l)$-qubit of noiseless communication in the diamond norm.  (This
is the usual requirement for transmission through noisy channel even
in the absence of adversarial attack.)  
We keep all the steps in QA+KG, except the $(m{+}l)$-qubit message
transmission is replaced by ${\cal N}'$.
In other words, we apply QA+KG to ${\cal N}'$ and additional
adversarial attack on ${\cal N}$ is transformed to an effective attack
on ${\cal N}'$.
Thus the security of this new protocol reduces to that of QA+KG.
Note that $l$ is sublinear in $m$ in QA, so this method has
communication rate similar to the case without an adversary, and does
not require any extra key.  In the end, regular error correction
handles the regular channel noise, and QA handles the remaining
adversarial attack.

A second way to see this solution is to note that PTP can be composed
with a mixed-state entanglement purification procotol (this solution
was informally suggested to us by Anne Broadbent).
For TQA', any composably secure ebits (with likewise secure classical
channel) gives a composably secure interactive quantum authentication
scheme.  There is no difference if these ebits are established via a
noisy channel and are purified, as long as the resulting ebits are
tested by PTP.
To obtain a noninteractive quantum authentication scheme, the
additional purification protocol has to be equivalent to an error
correcting code as described in \cite{bdsw96}, or \cite{BCGST02}, or
in step (2).

In similar spirits, the entanglement testing scheme in
\cite{aharonov2014local} can produce composably secure ebits using a
constant amount of secure quantum communication that only depends on
the desired accuracy.  This provides yet another secure authentication
scheme with slightly different initial resource requirement in the
interactive setting.  This new entanglement test can be further
adapted to an noninteractive by replacing the quantum communication by
trusted entanglement (in turns by using PTC with constant key size)
and authentication of classical messages (again of constant size).

Finally, when the transmitted state is known to the sender, the
lower bound for the key remains open.

%%%%%%%%%%%%%%%%%%%%%%%%%%%%%%%%%%%%%%%%%%%%%%%%%%%%%%%%%%%%%%%%%%%%%%%%
\paragraph{Acknowledgements}

We thank Daniel Gottesman for many crucial inputs to this
investigation, as well as Howard Barnum, Charles Bennett, Anne
Broadbent, Christopher Portmann, Jonathan Oppenheim, Louis Salvail,
Fang Song, and Henry Yuen for other interesting discussions.  In
particular, Charles Bennett suggested the application of quantum
authentication to QKD, and Anne Broadbent the main idea for the
solution for the noisy channel case.

This work was partly inspired by discussion at a randomization
workshop hosted by Claude Cr\'{e}peau at McGill's Bellairs Research
Station in 2003, and was made possible by the support of the IQI,
Caltech and the hospitality of the Perimeter Institute for all the
authors in August 2003.

This research has been supported in part by NSERC, CRC, CIFAR, ORF, 
QuantumWorks, the Croucher Foundation, the Tolman Foundation, the
Sherman Fairchild Foundation, AFOSR, the Simons Foundation, FQRNT, and
the NSF (grant no. EIA-0086038).

% \clearpage 
\appendix

% \input glossary.tex

% \begin{comment}

\section{Notation}

\label{app:glossary}

We gather notation used frequently in the paper, roughly in order
of first appearance:
\begin{itemize}

\item $\Phi$: A perfect EPR pair $\smfrac{1}{2} (|00\>+|11\>)(\<00|+\<11|)$
\item ebit: A unit for entanglement contained in $\Phi$

\item
QEnc: The particular encryption scheme that applies a random $m$-qubit
Pauli matrix $\vec \s_{xz}$ to an $m$-qubit message.
This requires $2m$ bits of key.

\item
QA: The particular noninteractive scheme proposed in \cite{BCGST02}
which in turn, applies QEnc, a quantum purity test code (PTC), and
then a secret syndrome.

\item PTC: A purity test code with error $\e$ is a set of quantum
codes such that given any nontrivial Pauli error, at least a fraction
$1-\e$ of the codes detect it.

\item PTP: A purity test protocol with error $\e$ is an LOCC scheme
that, with probability less than $\e$, outputs a quantum state
tagged \acc~but orthogonal to $m$ ebits.  

\item TP: Teleportation

\item
TQA': The interactive scheme to achieve quantum authentication using
TP, in which PTP is used to establish entanglement.

\item
TQA: a modification of TQA' in which Bob never tells Alice whether the 
entanglement is accepted or rejected. 

\item
KD$_\I$: In ideal key distribution box that simply provides Alice and
Bob a perfect, secret, shard key.  A variant was considered in
\cite{BHLMO-TCC05} that takes an auxiliary input bit, conditioned
on which either a key or an error message will be output.

\item
C$_\I$: A perfect classical channel, encrypted, authenticated, and hidden.  

\item
EBIT$_\I$: The ideal functionality for generating ebits, where an input 
{\sc acc} vs {\sc rej} will control whether the output is a number of 
perfect ebits or an error symbol.   

\item 
Q$_\I$: Our model of a perfect quantum channel which is TP[EBIT$_\I$,C$_\I$]. 

\item
QKD: Quantum key distribution

\item
QA+KG: QA augmented with recycling of the key used for QEnc if
authentication passes.  It is treated as a pair that performs QA and
key generation, while consuming a key from an ideal KD box KD$_\I$.

\item
Alice and Bob: Two honest parties trying to communicate
\item
Eve: An active adversary

\item
{\bf Capitalized letters often (though not always) denote random
  variables and the corresponding uncapitalized letters denote
  particular outcomes.}

\item
$\log$: Logarithm in base $2$

\item
$\mrho$: Generic symbol for a density matrix
\item
$|\cdot\>$: A vector in a Hilbert space, with label ``$\cdot$''
\\ $|\cdot\>\<\cdot|$: The projector onto the subspace spanned by
$|\cdot\>$, also known as ``outer-product'' of the ket $|\cdot\>$ and
the bra $\<\cdot|$.  We will simply write ``$\cdot$'' in place of the
bra and ket.
\item
$|\psi\>$: A unit vector or pure state.  Its density matrix is
given by $\psi = |\psi\>\<\psi|$.
\item
$\Tr(\cdot)$: The trace
\item
$\Tr_{\H_1}(\cdot)$: The partial trace over the system $\H_1$.  Let
$\mrho_{12}$ be the density matrix for a joint state on $\H_1$ and
$\H_2$.  $\Tr_{\H_1}(\mrho_{12})$ is the state after $\H_1$ is
discarded.
\item
$\| \cdot \|_1$: The Schatten 1-norm.  %The trace distance is half of it.
\item
$F$: The fidelity.  For two states $\mrho_1, \mrho_2$ in $H$,
$F(\mrho_1, \mrho_2) =
\max_{|\psi_1\>,|\psi_2\>} |\<\psi_1|\psi_2\>|^2$ where
$|\psi_{1,2}\> \in \H \ot \H'$ are
``purifications'' of $\mrho_{1,2}$ (i.e., $\Tr_{\H'}
|\psi_{1,2}\>\<\psi_{1,2}| = \mrho_{1,2}$), and $\<\cdot|\cdot\>$ is
the inner product.  Here, we can take $\dim(\H')=\dim(\H)$.
\item
$\s$, $\P$, $\s_\I$, $\P_\I$: $\s$ and $\P$ are generic labels for
protocols, with $\s$ possibly used as a subroutine.  The symbol of a
protocol with a subscript $\I$ denotes the ideal functionality of the
protocol.  \\
$\bullet$ $\P[\s]$: A protocol $\P$ calling a subroutine $\s$.  \\
$\bullet$ $\P_1$+$\P_2$: Conjoining two protocols $\P_1$ and $\P_2$. \\
Note that a subroutine $\s$ receives an input from the main protocol
$\P$ and returns an output to it.  In contrast, for conjoining
protocols, each may have its own input and output.  The protocols may
be run in parallel or sequentially.  In particular, information are
generally exchanged between the two, and each protocol may provide an
input to the other.\\
$\bullet$ $\P_1 = \P_2$ if they have the same circuit (but different 
interpretations and/or modular structures).
\item
$\CE$, $\S$: The environment and the simulator.  These are sets of
registers and operations and they are sometimes personified in our
discussion.
\item
$\G$, $\G_\I$: The random variables describing output bits of $\CE$
when interacting with $\P$ and $\P_\I+\S$ respectively.
\item
$\e$-\sr: $\P \; \e$-\sr$\P_\I$ is a shorthand for $\P$ $\e$-securely
realizes $\P_\I$ (see mathematical definition in \eq{usd}).  $\e$ is
called the {\em distinguishability-advantage} between $\P$ and
$\P_\I$.
\item
$T_\P$: The associated tree for a protocol $\P$

\end{itemize}

%----------------------------------------------------------------------
\section{The Simplified Universal Composability (Ben-Or-Mayers) model}
\label{app:qucmodel}

Our current setting is simpler than that considered in
\cite{QIP03-1,BM02} in two ways.  First, we are concerned with
unconditional security only.
Second, there is no unknown corruption of any party -- Alice and Bob
are honest and Eve is adversarial.  We do not use the formal
corruption rules.

We consider the acyclic quantum circuit model (see, for example,
\cite{Yao93a,AKN98}), with an important extension \cite{BM02} (see
also the endnotes \cite{poset}).  Throughout the paper, we only
consider circuits in the extended model. \\[0.5ex]
{\it 1. Structure of a protocol.} A (cryptographic) protocol $\P$ can
be viewed as a quantum circuit in the extended model
\cite{BM02,poset}, consisting of inputs, outputs, a set of registers,
and some partially ordered operations.

A protocol may consist of a number of subprotocols and parties.
Each subprotocol consists of smaller units called ``unit-roles,''
within which the operations are considered ``local.''
For example, the operations and registers of each party in each
subprotocol form a unit-role.
Communications between unit-roles within a subprotocol represent {\em
internal communications}; those between unit-roles in different
subprotocols represent input/output of data to the subprotocols.
A channel is modeled by an ordered pair of operations by the sender
and receiver on a shared register.
The channel available for the communication determines its security
features.\\
{\it 2. The game: security in terms of indistinguishability from the
ideal functionality.} Let $\P_\I$ denote the ideal functionality of
$\P$.  Intuitively, $\P$ is secure (in a sense defined by $\P_\I$) if
$\P$ and $\P_\I$ behave similarly under any adversarial attack.
``Similarity'' between $\P$ and $\P_\I$ is modeled by a game between
{\em an environment} $\CE$ and {\em a simulator} $\S$.  These are sets
of registers and operations to be defined, and they are sometimes
personified in our discussion.
In general, $\P$ and $\P_\I$ have very different internal structures
and are very distinguishable, and the simulator $\S$ is added to
$\P_\I$ to make an extended ideal protocol $\P_\I{+}\S$ that is less
distinguishable from $\P$.
$\CE$ consists of the adversaries that act against $\P$ and an
application protocol that calls $\P$ as a subprotocol.
At the beginning of the game, $\P$ or $\P_\I{+}\S$ are picked at
random.
$\CE$ will interact with the chosen protocol (running it and attacking
its vulnerabilities), and will output a bit $\G$ at the end of the
game.
The similarity between $\P$ and $\P_\I{+}\S$ (or the lack of it) is
captured in the statistical difference in the output bit $\G$.
See Fig.~1 for a summary of the game.
% See Figure~\ref{fig:compos} for a summary of the game.
\\[0.5ex]
{\it 3. Valid $\CE$.} The application and adversarial strategy of
$\CE$ are first chosen. (These cannot depend on whether $\CE$ is
interacting with $\P$ or $\P_\I{+}\S$.)
$\CE$ has to obey quantum mechanics, but is otherwise unlimited in
computation power.
If $\P$ is chosen in the game, $\CE$ can ({\sc i}) control the
input/output of $\P$, ({\sc ii}) attack insecure internal
communication as allowed by the channel type, ({\sc iii}) direct the
adversarial parties to interact with the honest parties in $\P$.
$\CE{+}\P$ has to be an acyclic circuit in the extended model
\cite{BM02,poset}.
Without loss of generality, an adversary can be modeled to only 
forward messages between the environment and the protocol, and 
the actual attack is executed by the environment.  (See Lemma 12
in \cite{Unruh09}.)  
\\[0.5ex]
{\it 4. Valid $\P_\I$ and $\S$.}
If $\P_\I+\S$ is chosen in the game, $\CE$ ({\sc i}) controls the
input/output of $\P_\I$ as before.
However, the interaction given by ({\sc ii}) and ({\sc iii}) above will
now occur between $\CE$ and $\S$ instead.  ($\S$ is impersonating or
simulating $\P$.)
The strategy of $\S$ can depend on the strategy of $\CE$.
$\P_\I$ should have the same input/output structure as $\P$, but is
otherwise arbitrary.  (Of course, the security definition is only
useful if $\P_\I$ carries the security features we want to prove for
$\P$.)
In particular, $\P_\I$ may be defined with internal channels and
adversaries different from those of $\P$.
$\S$ can ({\sc ii}$'$) attack insecure internal communication of $\P_\I$
and ({\sc iii}$'$) simulate the adversarial parties when interacting with the
honest parties in $\P_\I$.
Thus, $\P_\I$ exchanges information with $\S$, and this can modified
the security features of $\P_\I$.
To $\CE$, $\S$ acts like part of $\P_\I$, ``padding'' it to look
like $\P$, while to $\P_\I$, $\S$ acts like part of $\CE$.
It is amusing to think of $\S$ as making a ``man-in-the-middle'' attack
between $\CE$ and $\P_\I$.
Finally, $\CE{+}\P_\I{+}\S$ has to be an acyclic circuit in the
extended circuit model \cite{BM02,poset}.  Let the output bit be
$\G_\I$ in this case.
See Fig.~1 for a summary of the rules.
% See Figure~\ref{fig:compos} for a summary of the game.
%
% \vskip-3mm
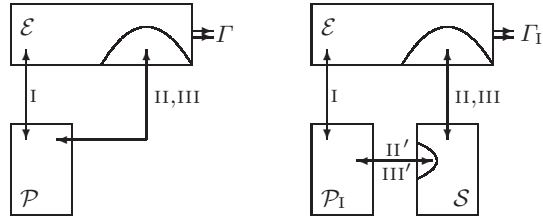
\begin{figure}[h]
\setlength{\unitlength}{0.4mm}
\centering
\begin{picture}(200,75)
% \put(0,0){\framebox(200,75){}}
\put(10,0){\framebox(20,30){}}
\put(13,3){\makebox{$\P$}}
\put(10,50){\framebox(60,20){}}
\put(70,59){\vector(1,0){7}}
\put(70,61){\vector(1,0){7}}
\put(76,55){\makebox(10,10){$\G$}}
\qbezier(40,50)(55,75)(70,50)
\put(55,25){\vector(-1,0){30}}
\put(55,25){\vector(0,1){30}}
\put(15,40){\vector(0,1){15}}
\put(15,40){\vector(0,-1){15}}
\put(13,60){\makebox{$\CE$}}
\put(60,36){\makebox(10,5){\sc ii,iii}}
\put(15,37){\makebox(5,5){\sc i}}

\put(115,37){\makebox(5,5){\sc i}}
\put(113,60){\makebox{$\CE$}}
\put(110,0){\framebox(20,30){}}
\put(113,3){\makebox{$\P_\I$}}
\put(110,50){\framebox(60,20){}}
\put(170,59){\vector(1,0){7}}
\put(170,61){\vector(1,0){7}}
\put(176,55){\makebox(14,10){$\G_\I$}}
\qbezier(140,50)(155,75)(170,50)
\put(155,35){\vector(0,1){20}}
\put(155,35){\vector(0,-1){10}}
\put(115,40){\vector(0,1){15}}
\put(115,40){\vector(0,-1){15}}
\put(145,0){\framebox(20,30){}}
\put(157,3){\makebox{$\S$}}
\qbezier(145,12)(158,18)(145,24)
\put(145,18){\vector(-1,0){20}}
\put(145,18){\vector(1,0){5}}
\put(132,19){\makebox(10,5){\sc ii}\hspace*{-0.5ex}$'$}
\put(132,11){\makebox(10,5){\sc iii}\hspace*{-0.3ex}$'$}
\put(160,36){\makebox(10,5){\sc ii,iii}}
\end{picture}
\label{fig:compos}
\caption{The game defining the composable security definition.  The
curved region in $\CE$ represents the adversaries against $\P$, and
the curved region in $\S$ represents the adversaries against $\P_\I$.
We label the types of interactions as described in the text.}
\end{figure}
% \vskip-3mm

\setcounter{theorem}{0}

We now restate the universal composable security definition and give a
slightly more extended description of the fundamental composability
theorems. \\[1ex]
{\bf Definition 1:} $\P$ is said to $\e$-securely realize $\P_\I$
(shorthand $\P \; \e$-\sr$\P_\I$) if
\be
    \forall \CE ~~\exists \S  ~~{\rm s.t.}~~
    \|\G-\G_\I \|_1 \leq \e
\;.
\label{eq:usd2}
\ee
We call $\e$ in \eq{usd2} the {\em distinguishability-advantage} between
$\P$ and $\P_\I$.
This security definition (in the model described) is useful because
security of basic composition follows ``by definition''
\cite{QIP03-1,BM02}.  We have the following simple version of a
universal composability theorem.
\begin{theorem} Suppose a protocol $\P$ calls a subroutine $\s$.
If $\s$ $\e_\s$-\sr$\s_\I$ and $\P[\s_\I]$ $\e_\P$-\sr$\P_\I$, then
$\P[\s]$ $\e$-\sr$\P_\I$ for $\e \leq \e_\P{+}\e_\s$.
\end{theorem}
Theorem 1 can be generalized to any arbitrary protocol with a proper
{\em modular structure}.  An example of an improper modular structure
is one with a security deadlock, in which the securities of two
components are interdependent.

Proper modular structures can be characterized as follows.
Let $\P[\s_1,\s_2{+},\cdots]$ be any arbitrary protocol using a
number of subprotocols.
This can be represented by a $1$-level tree, with $\P$ being the
parent and $\{\s_{i}\}$ the children.
For each $\s_{i}$ that uses other subprotocols, replace
the corresponding node by an appropriate $1$-level subtree.
This is done recursively, until the highest-level subprotocols (the
leaves) call no other subprotocols.  These are the primitives.
It was proved in \cite{BM02} that more general modular structures,
represented by an acyclic directed graph, can be transformed to a
tree.
The following composability theorem relates the security of a protocol
$\P$ to the security of all the components in the tree.
\begin{theorem} Let $\P$ be a protocol and $T_\P$ its associated
tree.
For each vertex $v$ in $T_\P$, let ${\cal M}_v$ be the subprotocol
corresponding to $v$ with its own subprotocols 
$\{{\cal N}_i\}_{i=1,\cdots,l}$.  (This can be an empty set if $v$ is
a leave.)
Then, if ${\cal M}_v[{\cal N}_{1\I},\cdots,{\cal N}_{l\I}]$
$\e_{{\cal M}_v}$-\sr${\cal M}_\I$, we have $\P$ $\e$-\sr$\P_\I$ for $\e
\leq \sum_{v} \e_{{\cal M}_v}$.
\end{theorem}
Theorem 2 is obtained by the recursive use of Theorem 1 and the
triangle inequality, replacing each subprotocol by its ideal
functionality, from the highest to the lowest level (from the 
leaves toward the root).
The distinguishability-advantage between $\P$ and $\P_\I$ is upper
bounded by the sum of all the individual distinguishability-advantages
between pairs of protocols before and after each replacement.

%----------------------------------------------------------------------
\section{The extended transpose trick}
\label{app:transposetrick}

One of the most useful tricks in quantum information
theory is that a transformation acting on one half of a
maximally entangled state can be implemented by applying a different
transformation acting on the other half.

We give an extended version of this trick allowing changes in the
dimensions.

Let $M = \sum_{j=1}^{d_2} \sum_{i=1}^{d_1} M_{ji} |j\>\<i|$ be a
possibly rectangular matrix, and $M^T = \sum_{j=1}^{d_2}
\sum_{i=1}^{d_1} M_{ji} |i\>\<j|$ be its transpose.

{\bf Lemma 1:}
$(M^T \otimes I) \sum_{j=1}^{d_2} |j\>|j\> =
 (I \otimes M) \sum_{i=1}^{d_1} |i\>|i\>$ ($\in
\mathbb{C}^{d_1}\otimes \mathbb{C}^{d_2}$).

{\bf Proof:} LHS
$= \sum_{j=1}^{d_2} (M^T|j\>)|j\> =
\sum_{j=1}^{d_2} \sum_{i=1}^{d_1} M_{ji} |i\>|j\> =
\sum_{i=1}^{d_1} |i\> (\sum_{j=1}^{d_2} M_{ji} |j\>)
= \sum_{i=1}^{d_1} |i\> (M|i\>)$ $=$
RHS.

Note that we are considering states differing from the maximally
entangled states by a {\em relevant} normalization.

{\bf Lemma 2:} let $U$ be a square matrix acting on systems $1$ and $2$ of
$d$ and $d_2$ dimensions.
Then,
\be
 (U_{12} \otimes I_3) \lbL |y\>_1 \sum_{j=1}^{d_2} |j\>_2 |j\>_3 \rbL =
 (I_{12} \otimes \<y|_4 U^T_{43}) \sum_{i=1}^{d d_2} |i\>_{12} |i\>_{43}
\,.
\ee

{\bf Proof:} The LHS can essentially be interpreted as the state obtained by
applying $U_{12} |y\>_1$ to system $2$, where $U_{12} |y\>_1$ is the
rectangular block of $U_{12}$ corresponding to $y$ ($d_2$ contiguous
columns).  Applying the first claim with $U_{12} |y\>_1 \ra M^T$ and
$d_1 \ra d d_2$, the resulting state is given by
$(I_{12} \otimes M_{43}) \sum_{i=1}^{d_1} |i\>_{12}|i\>_{43}$ 
exactly as claimed.

The LHS has the interpretation that we take $U_{12}$ as a real,
unitary matrix that encodes the logical state and the syndrome $y$
into the codeword acted on by error consistent with $y$.

The RHS, with $U_{12}^T = U_{12}^\dagger$ has the interpretation as a
decoding into the logical space and the syndrome, with {\em
postselection} on outcome $y$.

The equality in lemma 2 exactly proves the equivalence between
EBIT[PTC] and EBIT[PTP].

\section{Security of PSQA+KG}

\label{sec:psqa}

Recall from Section \ref{sec:qe} that in quantum encryption, Alice and
Bob share a key $K$ in which the realization $k$ occurs with
probability $p_k$.  To send a message $\mrho$, Alice transmits
$\CE_k(\mrho)$ and Bob applies $\CD_k$ to retrieve $\mrho$.  The
approximate soundness condition is given by
\[
\forall |\psi\>_{MR} ~~
\| (\CR\ot\CI)(\psi) - \rho_0 \otimes \Tr_M (\psi) \|_1 < \delta
\]
where
$\mrho_0$ is independent of $\psi$ and $\delta$ is a vanishing
security parameter.
We focus on $\CE_k(\mrho) = U_k \mrho U_k^\dagger$ and uniform $p_k$.

The particular protocol QEnc has $k = (x_1,z_1,\cdots,x_m,z_m)$ and
$U_k = \s_{x_1 z_1} \otimes \cdots \s_{x_m z_m} =: \s_{xz}$, 
with $2^{2m}$ values of $k$, or key size $2m$ bits.
In \cite{HLSW03}, another scheme called $\approx$QEnc is found
(existentially) such that only {\sc k} $=134 m 2^m/\delta^2$ values of
$k$ are used, but the unitaries $U_k$ are more complicated.
It has weaker security, in that $\forall \mrho$, $\|
\smfrac{1}{\mbox{\sc k}} \sum_k U_k \mrho U_k^\dagger -
\smfrac{I}{2^m}\|_\infty \leq \smfrac{\delta}{2^m}$, and it satisfies
the inequality of the soundness condition only if the adversary does
not have the purification of $\rho$ (a powerful form of quantum side
information), such as when $\rho$ is pure.

Recall from Section \ref{sec:qa} that the specific protocol QA given
in \cite{BCGST02} first applies QEnc to the $m$-qubit message,
followed by encoding with a purity test code (choose a code ${\cal
C}_t$ from a set based on a random value of $t$) and then applying an
operation corresponding to a random syndrome $y$ (all parameters as
described above).

Naturally, a question arises, whether one can replace QEnc by
$\approx$QEnc if the input for quantum message authentication is
promised to be pure.  We call the resulting protocol PSQA (standing
for Pure State QA), and again, we can append key recycling as an
additional step.

In this section, we prove the composable security of PSQA+KG.  We
believe the proof techniques are of independent interest.

The main challenge is to model the promise that a pure state is given
to Alice to be transmitted.  We handle this by imposing a restriction
on the environment, and call this restricted set of environments
$E^X$ in the analysis of PSQA+KG.  Fix a mapping between a set of
classical labels and a set of $m$-qubit pure quantum states $x \rightarrow
|\psi_x\>$.  The label can be real-valued and $|\psi_x\>$ unrestricted.
For each environment $\CE \in E^X$, $\CE$ can choose a value $x$ and
this results in Alice receiving an input $|\psi_x\>$ which is unknown to her.
One possible way this can happen is that a trusted party receives $x$
and then prepares $|\psi_x\>$ and gives it to Alice.  

Recall from Section \ref{sec:qucompos} that composable security can be
proved directly if for each environment $\CE$ (which is then fixed),
there exists a simulator, $\S$, such that PSQA+KG is indistinguishable from
an appropriately chosen ideal functionality conjoining $\S$.
For each $\CE$, this simulator $\S$ for PSQA+KG can be chosen to be the
simulator for a different protocol PSQA$^X$+KG (to be defined),
against the same environment $\CE$.
In PSQA$^X$+KG, an input $x$ is given to Alice, who prepares
$|\psi_x\>$ and then runs PSQA+KG.
Since the two protocols, PSQA+KG and PSQA$^X$+KG, are exactly
indistinguishable to each environment $\CE$, a simulator in the
analysis of latter gives the same distinguishability advantage for the
former.
It therefore suffices to prove composable security for PSQA$^X$+KG against
all $\CE \in E^X$.

The security proof for PSQA$^X$+KG is similar to that for QA+KG, and
can be done by defining a sequence of protocols, the first being
PSQA$^X$+KG and the last an ideal protocol, such that each protocol is
similar to the next.

The first protocol is PSQA$^X$+KG, the second is PSRQA$^X$+KG which is
the analogue of TQA+KG in which remote state preparation (RSP) is used
in place of teleportation, i.e., PSQA$^X$+KG =
(RSP+KG)[EBIT[PTC],C$_\I$].  PSRQA$^X$+KG uses ebits prepared by
insecure means.  The third protocol is (RSP+KG)[EBIT$_\I$,C$_\I$].  It
has a small probability of failure $\delta$ that is not caused by any
adversary, and provides an ideal functionality for the analysis.  The
analysis is identical to that for QA+KG, except for step (1).  It thus
remains to show the similarity between PSQA$^X$+KG and PSRQA$^X$+KG.

Recall from Section \ref{sec:qe} that there is a one-to-one
correspondence between $\approx$QEnc and RSP.  In RSP, Alice and Bob
shares $m$ ebits, and to transmit a state $\mrho$, Alice applies a
measurement ${\cal M}$ to her half of the ebits, and sends the outcome
to Bob.  The POVM has the form $\{\smfrac{1}{M}(U_k \mrho
U_k^\dagger)^T\}_k$ for $M=\| \sum_k U_k \mrho U_k^\dagger \|_\infty$,
together with an extra POVM element $F = I-\smfrac{1}{M} (\sum_k U_k
\mrho U_k^\dagger)^T$ (we say that $k=f$).  Conditioned on an outcome
$k \neq f$, Bob's half of the ebits becomes $U_k \mrho U_k^\dagger$
(this can be proved by the transpose trick in the previous appendix),
but if $k=f$, RSP fails.  Ref. \cite{HLSW03} shows that taking
$134 m 2^{m}/\delta^2$ different $U_k$'s is sufficient to ensure that
$\Pr(f) < \delta$.

Once again, consider schematic diagrams, now for PSQA$^X$+KG and
PSRQA$^X$+KG:\\
\bea
\epsfig{file=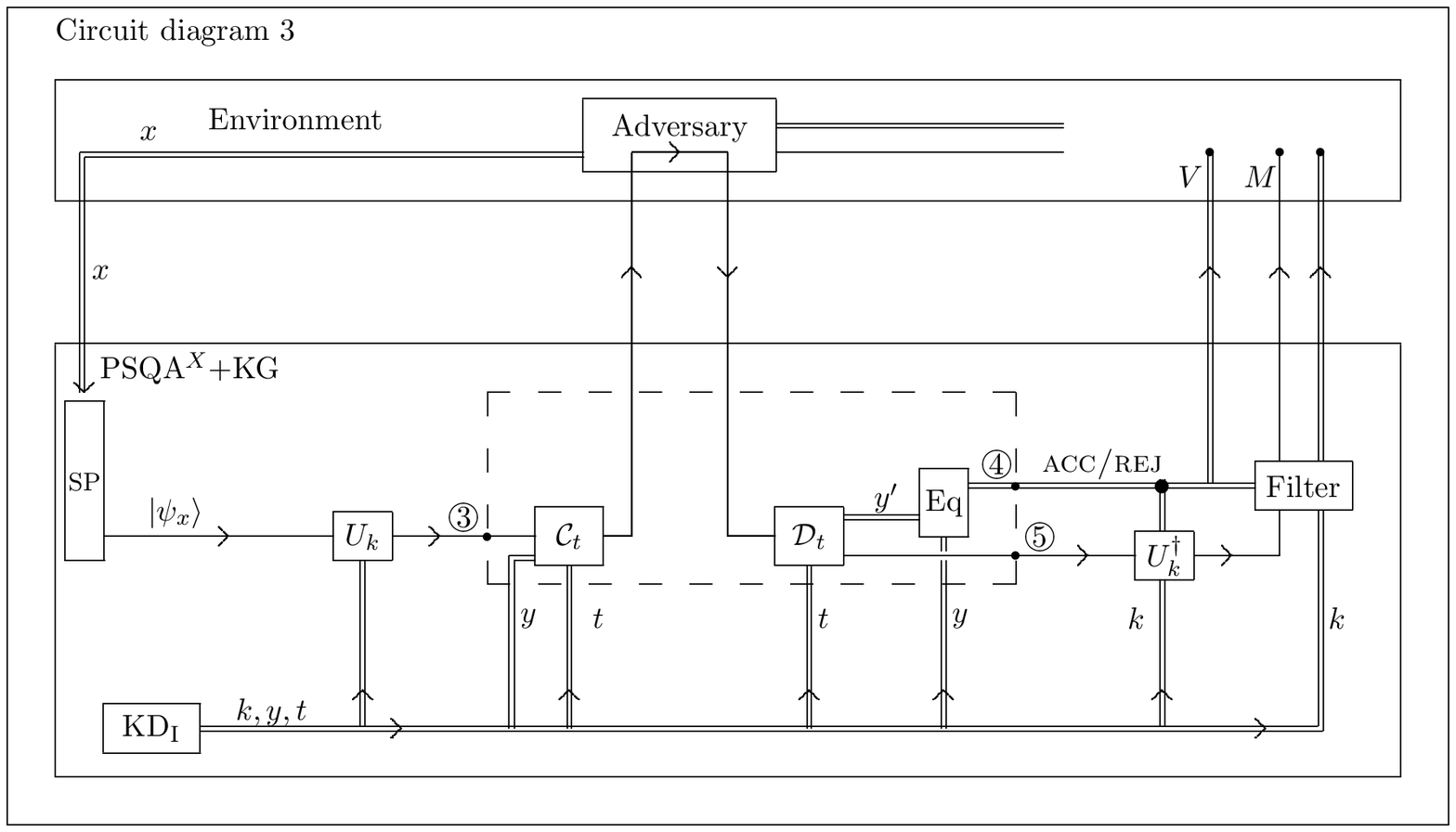,width=6.0in}
\nonumber
\eea
\bea
\epsfig{file=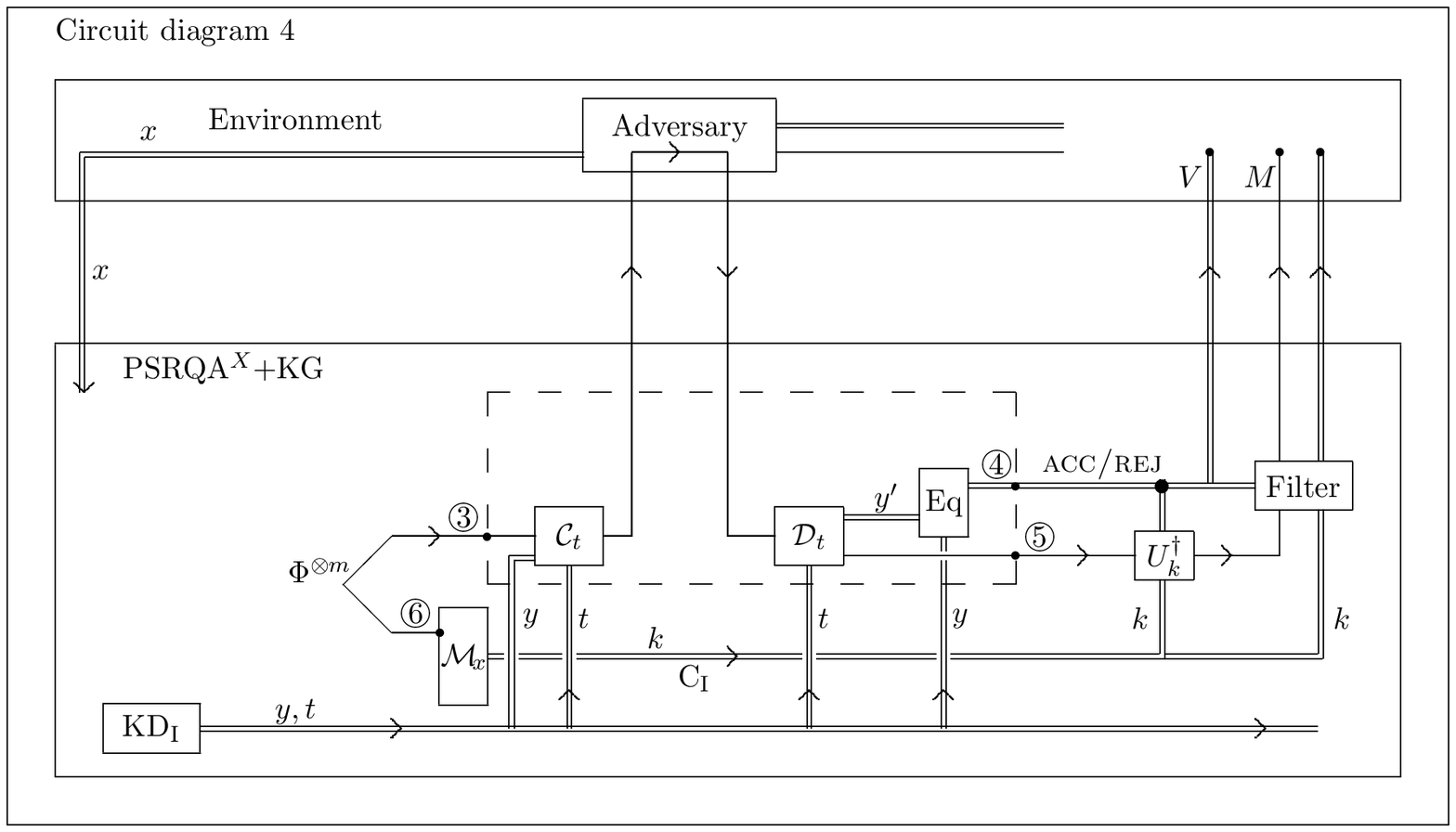,width=6.0in}
\nonumber
\eea

In PSRQA$^X$+KG, Alice's measurement ${\cal M}$ is defined with $\mrho
= \psi_x$ and this requires her knowledge of $x$.  This is why we
focus on PSQA$^X$+KG and use its composable security to infer that of
PSQA+KG.  If $k \neq f$, the states label \textcircled{3} in the two
protocol are identical.
This introduces an additional contribution of $\delta$ to the 
distinguishability advantage.  
The rest of the analysis is identical to that of QA+KG.  
An additional $\delta$ results from the difference between
RSP[EBIT$_\I$,C$_\I$] and Q$_\I$.  

Putting everything together, PSQA+KG $(2 \sqrt{2} \e^{1/3}+2\delta)$-s.r.
Q$_\I$+KD$_\I$.

\section{Quantum universal composable security of the Wegman-Carter
scheme}

\label{sec:wc}

We consider a Wegman-Carter type of authentication scheme (WC) that
does the following.  Let ${\cal H}= \{h_k\}_k$ be an
$\e$-almost-strongly universal$_2$ family of hash functions from the
set of messages ${\cal M}$ to the set of authentication tags ${\cal
T}$.  Let $k,t$ be the value of the shared key.  If the message is
$x$, Alice transmits $(x,h_k(x) \oplus t)$ to Bob, where $\oplus$
represents the bitwise {\sc xor}.  In other words, the hash value of
the message is one-time-padded with the key $t$.  Usually, when
executing this scheme WC, the same hash function is reused for
subsequent messages.

Here, we analyze WC+KG, which runs WC and recycles the key $k$, in a
way similar to but much simpler than QA+KG.  In particular, we let
$k,t$ be the output from an internal KD$_\I$ box that provides the
keys to WC.  Note that recycling of the key $k$ is more general than
reusing of the hash function (universal vs non-universal
composability).

The ideal functionality has two parts.  The first part is a magic
authentication box that sends the message $x$ via an insecure channel,
and the received message is $x'$.  The output is ($x$, \acc) if
$x=x'$, and (\rej) if $x \neq x'$.  The second part is KD$_\I$ which
outputs a perfect key $k$ between Alice and Bob.  

The simulator runs the ideal authentication box to transmit $x$ to the
environment, but appends a random tag $h$, and receives $x',h'$.  The
simulator checks if $h=h'$ and feeds $x'$ into the ideal
authentication box, which then outputs ($x$,\acc) or (\rej).  If
$t=t'$ and the output is ($x$,\acc), the simulator makes the final
output ($x$,$t$,\acc,$k$), else, the simulator outputs (\rej,$k$).

WC+KG and the ideal functionality differ only if $(x,h) \neq (x',h')$
and the former accepts.  This happens with probability less than
$\e$, which then upper bounds the distinguishability
advantage.

Due to \cite{P13}, we are now aware of a problem that in principle, an
adversary can guess a key value and tamper with the message
accordingly, and the subsequent \acc~or \rej~output leaks information
on the key to be recycled.  This introduces an additional contribution
to the distinguishability advantage.  We leave the work of correcting
the above proof for a later version of this manuscript.

% \bibliographystyle{unsrt}
% \bibliography{qauth}

\begin{thebibliography}{10}
\bibitem{qasecdef} In \cite{AMTW00} and \cite{BCGST02}, the security
  definition and security proofs are given without explicitly
  considering the purifying system. When exact security definitions
  are achievable (as in \cite{AMTW00}), the former implies the latter
  by linearity. However, \cite{HLSW03} shows that an approximate
  condition holding for all pure states need not imply the same
  condition in the presence of purifications. In particular, for
  quantum encryption, security in the presence of purifications
  doubles the key size.  For the QA protocol in \cite{BCGST02}, the
  security is proved explicitly for the security definition without
  the purification, though it is not difficult to check that QA
  satisfies the stronger security definition with purifications. (This
  also follows immediately from our proofs.) Meanwhile, QA uses QEnc
  as a subroutine, and the lower bound for the keysize comes from the
  most general quantum encryption. It follows from \cite{HLSW03} that
  the correct lower bound is $m$ bits for pure state schemes and $2m$
  bits in the presence of purifications, instead of the $2m$ bits
  stated in \cite{BCGST02}). Appendix \ref{sec:psqa} shows that the
  lower bound for pure state is attainable asymptotically. To keep our
  discussion simple, we stick to the stronger security definitions
  including purifications, except for in Appendix \ref{sec:psqa}.
\end{thebibliography}

\end{document}